\documentclass[twocolumn,fleqn,usenatbib]{mnras}

\usepackage{graphicx}
\usepackage{color}
\usepackage[nointegrals]{wasysym}
\usepackage{mathtools}
\usepackage{subcaption}

\usepackage{newtxtext,newtxmath}
\usepackage[T1]{fontenc}
\usepackage{bm}
\usepackage{amsmath}
\renewcommand{\vec}[1]{\ensuremath{\bm{{#1}}}} 
 
\newcommand{\diff}{\ensuremath{\mathrm{d}}}
\newcommand{\deriv}[2]{\ensuremath{\frac{\diff #1}{\diff #2}}}
\newcommand{\pderiv}[2]{\ensuremath{\frac{\partial #1}{\partial #2}}}

\DeclareRobustCommand{\VAN}[3]{#2}
\let\VANthebibliography\thebibliography
\def\thebibliography{\DeclareRobustCommand{\VAN}[3]{##3}\VANthebibliography}

\title[A well-balanced multi-purpose hydrodynamic code for planetary atmospheres]{\textsc{Aiolos} - A multi-purpose 1-D hydrodynamics code for planetary atmospheres}

\author[M. Schulik and R. Booth]{
Matthaeus Schulik,$^{1}$\thanks{E-mail: mschulik@ic.ac.uk (Imperial)}
Richard A. Booth,$^{1}$
\\
$^{1}$Imperial College London, Kensington Lane, London W1J 0BQ, UK}

\date{Accepted XXX. Received YYY; in original form ZZZ}

\pubyear{2022}

\begin{document}

\label{firstpage}
\pagerange{\pageref{firstpage}--\pageref{lastpage}}
\maketitle

\begin{abstract}
We present a new 1-D multi-physics simulation code with use cases intended for, but not limited to, hydrodynamic escape problems of planetary atmospheres and planetary accretion models. Our formulation treats an arbitrary number of species as separated hydrodynamic fields, couples them via friction laws, allows for a multi-band flux-limited radiation transport, and tracks ionization fronts in high-energy irradiation bands. Besides coupling various known numerical solution techniques together, we improve on the numerical stability of deep hydrostatic atmospheres by using a well-balanced scheme, hence preventing unphysical driving of atmospheric in- or outflow. 
We demonstrate the correct physical behaviour of the individual code modules and present a few simple, new applications, such as a proof-of-concept simulations of combined core-powered mass-loss and UV-driven atmospheric escape, along with a fully time-dependent core-collapse giant planet simulation.
The multi-species nature of the code opens up the area of exploring simulations that are agnostic towards the dominant atmospheric species and can lead to implementations of advanced planetary evolution schemes.
\end{abstract}

%

\section{Introduction}

The exoplanet population as it presents itself to the scientific community today exhibits a number of striking features. Those features manifest themselves in the data as gaps in the period-radius and period-mass distributions that are upheld after bias corrections \citep{mayor2011, Fulton2017, vaneylen2018}.
The period-radius gap at small planetary radii of $\rm \sim1.7 R_{\oplus}$ is often interpreted as a sign of past and efficient hydrodynamic escape \citep{Owen2017, Ginzburg2018}, an imprint of the formation semimajor-axis in the protoplanetary disc \citep{leeconnors2021, lee2022} or of compositional differences of the planetary bulk \citep{zeng2019}. Furthermore, the paucity in intermediate-radius planets at high planetary equilibrium temperatures \citep{mayor2011} is often interpreted as a relic of unsuccessful runaway gas-accretion \citep{mordasini2014} owing to an exceedingly long phase of quasi-hydrostatic gas-envelope growth around massive cores \citep{pollack1996} under high-opacity disc conditions \citep{movshovitz2010}.

Both of these mechanisms, gas accretion \citep{mizuno1978, pollack1986, wuchterl1990, pisoyoudin2014} and hydrodynamic escape \citep{opik1963, hunten1973, sekiya1980, watson1981, zahnlekasting1986, yelle2004, lammer2008}, have a long history of being studied theoretically under various assumptions, going back to the works of \cite{bondi1952, parker1958}. While various modern codes exist to simulate those processes in dynamical multi-dimensional simulations \citep{kley1998, ayliffe2009a, dangelo2003, tanigawa2012, vidal-madjar2004, mccann2019}, however these codes make simplifications to the computed physics due to their enormous computational cost. We prioritize developing the multi-physics aspect of a 1-D code in this work, as we ultimately aim for evolutionary simulations covering significant multiples of dynamical atmospheric timescales. Furthermore we aim to follow the evolution of the planetary atmospheric composition through in- and outflows. 

Particularly when considering that in the context of planet formation and evolution, accretion and escape processes act subsequently, it is desirable to write a simulation code which can compute both situations reliably, while simultaneously expanding upon past work. 

From the standpoint of numerical simulations, such a code therefore should be able to capture transonic phenomena, handle optically thin-to-thick transitions and reproduce hydrostatic equilibria in a numerically stable manner. All the latter properties are required in a dynamical simulation, as the states of atmospheric accretion and escape can be understood as nonlinear perturbations on top of an initially hydrostatic state, where only the sign of the pressure perturbation differs. 

The sign of this perturbation depends on whether heating or cooling is more important in a given atmospheric region. While accretion will be driven by bolometric cooling and self-gravity\citep{hubeny1990, pollack1996}, atmospheric escape can be powered both by bolometric \citep{Ginzburg2018, gupta2019, gupta2020} and high-energy irradiation \citep{sekiya1980,Lammer2003, yelle2004, garciamunoz2007, MurrayClay2009,Owen2012, erkaev2013, kubyshkina2018, kubyshkina2019}.
Recent studies have particularly targeted the relative importance of the latter two processes for planetary evolution \citep{RogersOwen2020, rogers2021}, but ab-initio simulations of bolometrically driven mass-loss (i.e. core-powere mass-loss) and a unified bolometric-UV simulation framework have not been conducted. 

In the numerical context, traditional codes lack some important aspects which are desired to solve a hydrodynamic multi-species problem with bolometric and high-energy radiation transport, hence we now briefly outline our technical reasoning for presenting a new simulation code.

Firstly, and most importantly, modern Riemann solvers suffer from problems keeping a hydrostatic solution stable against gravity source terms, a problem already described in \cite{greenberg1996}. A naive approach in trying to balance a fluid will induce oscillations that result in catastrophic imbalances \citep{Kappeli2016} because the cell-centered gravitational source does not `know' about the pressure differences that were passed to the interface-centered Riemann fluxes, and therefore cannot exactly fulfill a hydrostatic balance condition against them. An alternative for this is to include the gravitational flux in the Riemann solver \citep{colella1984}, which would require a new numerical flux function and an efficient and correct approximation to this numerical flux, which is not a generally trivial task \citep{Toro1994}. 

To get around those problems, one needs to \textit{well-balance} sources and flux differences in the cell in order to generate zero net momentum flux in cases when the forces are balanced out. A possible approach, which we employ throughout this work, is given in e.g. \cite{kappeli2014, Kappeli2016}, where a reconstruction approach is used to correct the pressure at cell interfaces with information about the force balance which needs to be fulfilled.
This allows a time-dependent simulation to remain stable against oscillations as well as increasing the fidelity of hydrodynamic states in the deep, near-hydrostatic atmosphere.

Furthermore, in moving fluid regions, the tracking of velocities and temperatures of individual particle species is important \citep{schunk1977} in order to determine whether they escape or remain hydrostatic \citep{zahnlekasting1986}.
As particles interact with each other collisionally, it is required to solve for the rates of their momentum and internal energy exchange as well. Only then accurate velocities and escape/accretion rates for all species can be obtained.

Single-fluid formulations often avoid solving this problem by evolving an individual species density in a diffusion approximation, while a single velocity average and temperature variable exists for all species. Those formulations can be problematic and unflexible however, and lead can lead to inaccurate low-density behaviour already in two-species solutions \citep{zahnlekasting1986}. The process of decoupling is important in upper planetary atmospheres \citep{schunk1980}, where the (partially ionized) plasma species can develop their own temperatures and velocities. In reality, this process is often aided by magnetic fields \citep{lammer2008, gunell2018}, which we entirely neglect in our work, but monemtum or thermal decoupling can still play a major role for the atmospheric evolution. We use a recently presented method by \cite{benitez-llambay2019} for linearized drag-laws that allows the frictional coupling of an arbitrary number of species. We employ their formulation and develop a consistent formulation for the heat exchange terms into our temperature solver, which allows us to correctly represent any coupling strength.

Effects of radiation transport on atmospheric outflows are often treated in the popular energy-limited escape approximation \citep{watson1981, Lammer2003, baraffe2004, owen2019}. This formulation neglects radiative losses completely, or re-introduces them as non-thermal cooling functions \cite{MurrayClay2009, johnstone2018, kubyshkina2018, kubyshkina2019} without ray-tracing. The escape rate is then given by balancing the cooling-corrected stellar high-energy heating with the adiabatic expansion of the atmosphere into space. 
Selecting the high-energy parts of the stellar radiation spectrum for an energy-limited escape treatment while neglecting the thermal photon field is justified by the inefficient thermalization of atomic species with the ambient thermal photon field \citep{sekiya1980, erkaev2007, owenalvarez2016, salz2016, kubyshkina2018}.
Molecular species on the other hand have dense line forests, making them efficient at losing energy towards the ambient thermal photon field. Those species can then develop complex thermal profiles \citep{guillot2010, parmentier2014, parmentier2015}. The effects of those thermal profiles on atmospheric escape have not yet been characterized, which is required in order to properly characterize core-powered mass-loss.
It is therefore desirable to obtain a code that can erase the artificial dividing line between the heating/cooling physics of molecules and atoms, as well as solve for the transition between them via photodissociation and recombination.

To this end, we present \textsc{aiolos}, a new public
\footnote{The code will be accessible on github after the refereeing and cleanup process is finished.}
, time-dependent 1-D multi-species, multi-band, radiation-hydrodynamics code. In Section 2 we first present the physical equations solved. In Section 3 we present and discuss the numerical schemes used for hydrostatic well-balancing, drag, radiation transport and photochemistry. Section 4-7 shows those the performance of those individual physics modules compared against analytical solutions or numerically solutions from the literature. Section 8 presents two new applications that our approach enables. In Section 9 we finally discuss limitations in our approach and some important caveats.

\section{Equations solved}
\label{sec:equations}

\textsc{Aiolos} solves the equations of multi-species radiation-hydrodynamics in one dimension, in either cartesian, cylindrical or spherical coordinates. All results discussed in this paper are computed in spherical coordinates, as is usual for planetary appliciations, except for shock tube tests, which are performed in cartesian geometry. Each species is treated by solving an independent set of Euler equations. The different species coupled together via collisions, radiative heating and cooling and photo-chemistry. Explicitly, for each species, $s$, the density, $\rho_s$, momentum, $\rho_s u_s$, and total energy, $E_s = \tfrac{1}{2}\rho_s u_s^2 + \rho_s e_s$, obey 
\begin{align}
    \pderiv{\rho_s}{t} + \nabla \cdot [\rho_s u_s] &= Q_s, \label{eqn:hydro_mass} \\
    \pderiv{\rho_s u_s}{t} + \nabla \cdot [\rho_s u_s^2 + p_s] &= - \rho_s \nabla \Phi + \deriv{\mathcal{M}_s}{t} + Q'_s, \\
    \pderiv{E_s}{t} + \nabla \cdot [u_s (E_s + p_s)] &= - u_s \rho_s \nabla \Phi + u_s \deriv{\mathcal{M}_s}{t} + \deriv{\mathcal{E}_s}{t} + \Gamma_s. \label{eqn:hydro_en}
\end{align}
Here $p_s$ is the partial pressure of the species, $e_s$ is the internal energy per unit volume and $\Phi$ is the gravitational potential. We close these equations with the adiabatic equation of state $p_s = \left(\gamma_s-1\right) \rho_s e_s$ and ideal gas equations of state for the specific internal energy, i.e. $e_s = c_{v,s} T_s$. The adiabatic index $\gamma_s$ is computed from the species degrees of freedom  $f_s$ via $\gamma_s = (f_s+2)/f_s$ and the constant-volume heat capacity $c_{v,s}$ is computed from the maximum atomic heat capacity of hydrogen $R_{\rm H}\approx 8.3144\times 10^7\rm \,erg\,K^{-1}\,mole^{-1}$ via $c_{v,s} = \frac{1}{2} f_s R_{\rm H} / \mu_s$, where $\mu_s$ is the mass of the species in $amu=1.66054\times10^{-24}$g. We keep $\gamma_s$ and $c_{v,s}$ constant for all species throughout the simulations presented in this work.
Furthermore, $Q_s$ and $Q'_s$ encode the sources and sinks of mass and momentum for each species, while $\Gamma_s$ encodes the heating and cooling. The exchange of momentum and energy due to collisions are given by \citep{schunk1977, schunk1980}
\begin{align}
    \deriv{\mathcal{M}_s}{t} &= \rho_s \sum_{s'} \alpha_{ss'} (u_{s'} - u_{s}) \label{eqn:coll_mom} \\
    \deriv{\mathcal{E}_s}{t} &= \rho_s \sum_{s'} \frac{\alpha_{ss'}}{m_s + m_{s'}} \left[ m_{s'} (u_{s'} - u_{s})^2 + 3 k_{\rm B} (T_{s'} - T_{s}) \right]. \label{eqn:coll_en}
\end{align}
Here $m_s$ is the mass of the species $s$, the species temperature is $T_s$ and $k_B$ is the Boltzmann constant. Momentum conservation requires that the collision frequencies, $\alpha_{ss'}$, obey $\rho_s \alpha_{ss'} = \rho_{s'} \alpha_{s's}$. The first term in \autoref{eqn:coll_en} is simply the energy generated by drag heating, while the second is the internal energy exchanged in collisions. We note that the drag terms conserve energy, i.e. $\sum_s u_s \deriv{\mathcal{M}_s}{t} + \deriv{\mathcal{E}_s}{t} = 0$.

Optionally, we couple these equations with the equations of radiation transport. \textsc{Aiolos} treats radiation transport using a multi-band version \citep{vaytet2013a} of the hybrid flux-limited diffusion approximation \citep{kuiper2010, commercon2011, bitsch2013a}. In practice this means that the incoming and potentially ionising solar irradiation is solved for by direct ray-tracing in plane-parallel geometry, whereas the exchange of thermal energy between gas and photons is solved via the flux-limited diffusion approximation \citep{levermore1981} in spherical geometry. When the radiation module is activated, we compute
\begin{equation}
    \Gamma_s = \sum_{b}^{\rm bands} \rho_s \kappa^b_{\mathrm{P},s} \left[4\upi J^b - f^b(T_s) 4 \sigma_{SB} T_s^4\right] + \Delta S_{\odot, s} + \Lambda_{s},
\end{equation}
where $\sigma_{SB}=5.67\times10^{-5}\,\rm erg\,s^{-1}\,cm^{-2}\,K^{-4}$ is the Stefan-Boltzmann radiation constant,  $J^b$ is mean photon intensity in the band, $b$,  $\kappa^b_{\mathrm{P},s}$ is the Planck-mean opacity of species $s$ averaged over the band $b$ and $f^b(T_s)$ is the fraction of the total black-body radiation at $T_s$ that is emitted into each band.
In the hybrid method, which is the default setting, the direct heating from the stellar radiation, $\Delta S_{\odot,s}$ and high-energy cooling as well as non-thermal cooling from collisions with free electrons $\Lambda_s$(which we call high-energy cooling), is included in the internal energy equation as stated above. This approach is chosen rather than solving a two-stream system for the radiation. Since \textsc{aiolos} is a 1D code, we treat the stellar irradiation in the surface-averaged approximation \citep[e.g.][]{guillot2010}, where the exact local expression becomes:
\begin{align} 
 \Delta S_{\odot,s} &= \frac{1}{4}\int \frac{ F_{\odot}(\nu)}{h\nu}  \exp(-\tau_{\odot}(\nu)) \rho_s \kappa_{\odot,s}(\nu) \diff \nu \label{eq:stellar_heating_0}  \\  
    &\approx \frac{1}{4} \sum_{b}^{\rm bands} F_{\odot}^{b}  \exp(-\tau_{\odot}^{b}) \rho_s \kappa_{\odot,s}^{b} \label{eq:stellar_heating_1}.
\end{align}
where the factor $1/4$ results from the ratio of geometric planetary absorption cross-section to total planetary surface, over which the absorbed energy is distributed. Furthermore, $\tau_{\odot}^b$ is the total optical depth to the stellar radiation computed from the Planck-mean opacity to solar radiation in the b-band $\kappa_{\odot,s}^b$ and $F_{\odot}^b$ is the stellar flux at $\tau^b=0$, and the energies of stellar photons at frequency $\nu$ are $h\nu$. In practice, the frequency-dependence is absorbed into a finite number of bands. In the cases where ionizing radiation can change the density $\rho_s$, the optical depth and ionization rate have to be solved for simultaneously, we describe this process in Sect. \ref{sec:numerical_ion}. The mean intensity of the photon field, $J^b$, obeys the time-dependent equation
\begin{equation}
    \frac{1}{c}\pderiv{J^b}{t} + \frac{1}{4\upi}\nabla \cdot F^b = - \sum_{s}^{\rm species} \rho_s \kappa^b_s \left[J^b - f^b(T_s) \frac{\sigma_{SB}}{\upi} T_s^4\right] \label{eqn:radiation_equation}
\end{equation}
where the photon flux $F^b$ is computed in the Flux-Limited Diffusion approximation:
\begin{equation}
    F^b = - \frac{4\upi \lambda(J^b)} {\sum_s\rho_s \kappa^b_s} \nabla J^b.
    \label{eqn:radtrans_FLD_flux}
\end{equation}
For the  flux limiter, $\lambda(J_b)$ \citep{levermore1981}, we use the version given by \citet{kley1989}. 
\begin{align}
    \lambda(J) \equiv \lambda(R) = \begin{cases}
\frac{2}{3+\sqrt{9+10R^2}} & R \leq 2\\
\frac{10}{10R+9+\sqrt{81+180R}} & R > 2
\end{cases} \label{eq:radtrans_fluxlimiter}
\end{align}
where the photon mean-free path parameter, $R$, is
\begin{align}
   R=\frac{\xi}{\sum_s\rho_s \kappa^b_s} \frac{|\nabla J|}{J}.
   \label{eq:radtrans_r_parameter}
\end{align}

The purpose of the flux limiter is to regularize the flux in optically thin regions, where without it the diffusion approximation would produce radiation that travels faster than the speed of light. With the flux limiter included the maximum flux obtained is $|F^b| = 4\upi J/\xi = cE_{\rm R}/\xi$, where $E_{\rm R}$ is the energy density of radiation. In other words, for the standard choice, $\xi=1$, the maximum speed at which the radiation energy is transported is the speed of light. The parameter $\xi=1$ is however just a choice, and any value of $\xi \ge 1$ would obey causality while also producing the correct behaviour in the optically thick regime (where the flux is independent of $R$).

As we will show later, $\xi=1$ may not be the best choice for planet evaporation models. The reason behind this is that the optical depth can vary rapidly on length scales smaller than the planet's radius. In such a case the atmosphere can be well modeled by a plane-parallel slab, for which $H=F/4\upi= J/2$ is a good approximation for the flux in the optically thin regions \citep[e.g.][]{guillot2010}. Hence $\xi=2$ seems natural for planetary atmosphere applications. That $\xi > 1$ holds in a plane-parallel atmosphere is evident from the fact there is always a significant flux of radiation in the close-to horizontal directions so the average flux of radiation away from the planet must be less than the maximal case, in which all of the radiation is travelling directly away from the planet. However, far from the planet the radiation will seem as if it is emitted from a point source, and in such a case it is straight forward to verify that $\xi=1$ is appropriate. This makes it clear that $\xi$ (or more directly, the Eddington tensor) must vary with radius. Although this can be accounted for, e.g. in a variable Eddington tensor method, we simply note that while $\xi = 1$ may be favourable for some problems it is sensible to allow users to choose the value of $\xi$ that best suits their problem. We highlight the physical effects of the choice of $\xi$ in Sect. \ref{sec:results_rad}.



\begin{figure*}
\hspace*{-0.5cm}        
  \centering
   \includegraphics[width=0.35\textwidth]{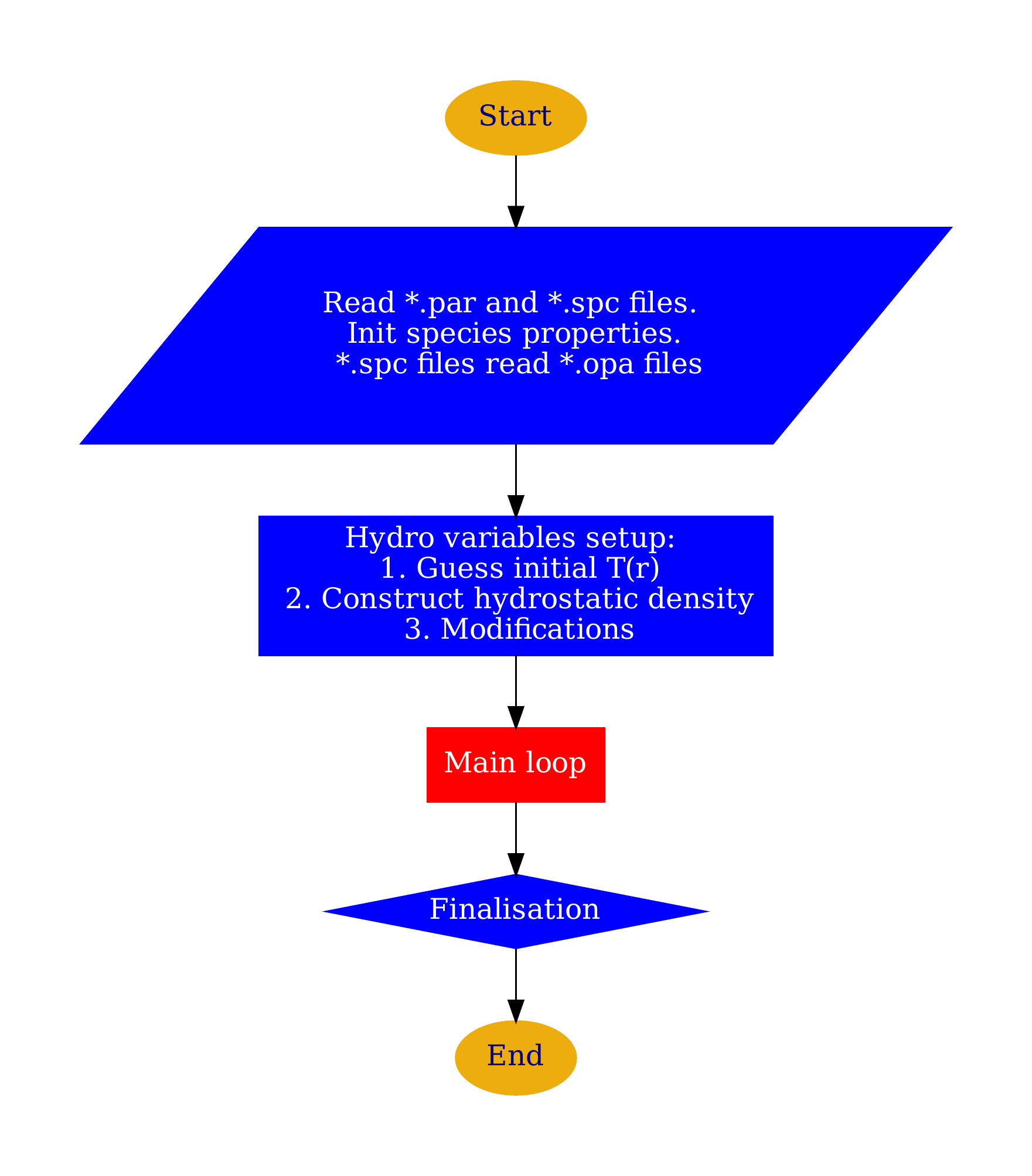}
   \includegraphics[width=0.4\textwidth]{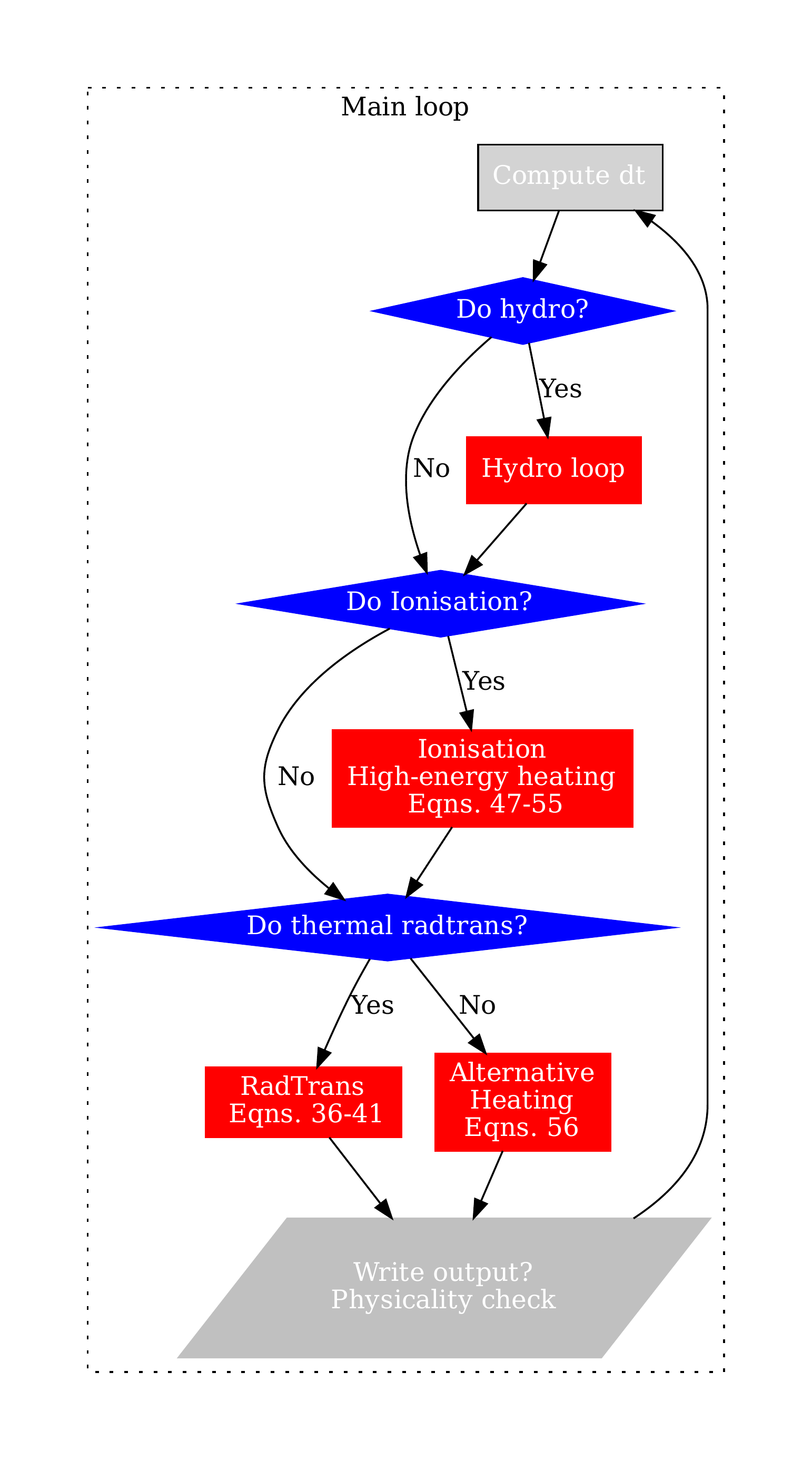}
   \includegraphics[width=0.25\textwidth]{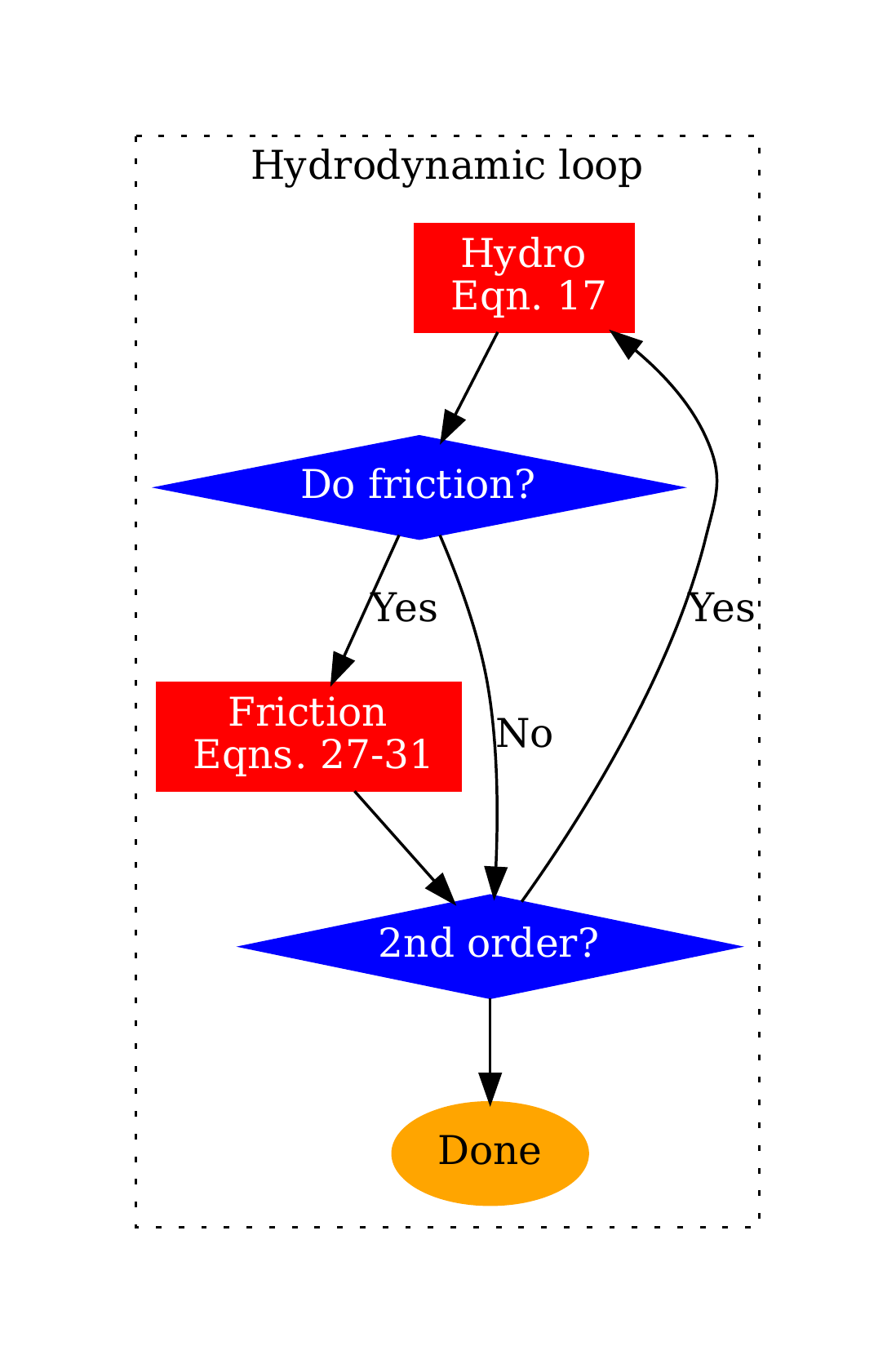}
\caption{Flow charts to illustrate the data flow in \textsc{AIOLOS}. $\rm *.par$ files contain global simulation parameters, $\rm *.spc$ files are a list of all species properties, such as $m_s$, $f_s$, relative initial amount, further links to opacitiy files etc. All modules (hydrodynamic, friction, ionisation, radiation transport) can be switched on and off via the $\rm *par$-file, allowing for different physical approximations to be realised in the main program loop, such as hydrostatic radiation transport or hydrodynamic isothermal runs.} 
\label{fig:flowchart}

\vspace*{+0.20cm}
\end{figure*}

\section{Numerical methods}
\label{sec:numerical}

We solve equations~(\ref{eqn:hydro_mass})--(\ref{eqn:hydro_en}) and (\ref{eqn:radiation_equation}) using a finite-volume approach. The spatial domain is discretized into a series of cells on a fixed grid. Each cell, with the numbering $i$ starting at 0, is specified by its cell centre, $r_i$, and edges, $r_{i-1/2}$ and $r_{i+1/2}$, which have corresponding surface areas $A_{i-1/2}$ and volumes $V_i$. We denote by $\vec{U}^n_{s,i} = \{\rho_{s,i}, \rho_{s,i}u_{s,i}, E_{s,i}\}$ and $J^n_{b,i}$ the average of the conserved fluid variables and mean-intensity of radiation over the volume of cell $i$ at time $t_n$, respectively. 
Fig. \ref{fig:flowchart} illustrates the work flow of the program in a global overview. Details on the individual substeps are presented in the following subsections. All substeps are solved via simple operator-splitting \citep{stonenorman1992} in order to construct a full solution after one timestep. As shown in Fig. \ref{fig:flowchart}, the code first reads in the initial parameters in order to construct the initial conditions $\vec{U}^n_{s,i}$ per species. This requires first, a guess of the temperature profile, adiabatic or isothermal by default, from which a hydrostatic density profile is constructed in the given gravitational potential. The code then begins the main loop, evolving the system of equations. This consists of the hydrodynamic subloop, which evolves each species separately first and then secondly couples their velocities, in case the friction flag is set. 
After the hydrodynamic part, first the opacities are calculated per species, based on the given density and temperature. This results in knowledge of the optical depths. In the case that the incoming radiation is ionising, optical depth and ionisation rates need to be self-consistently computed in the ionization routine. Once the optical depths are self-consistent, the heating function can be computed from the optical depth per cell. With the heating function per cell known, the radiation transport step is finally solved, updating all species' temperatures and the thermal radiation fields. The code performs a positivity check of important quantities such as $\rho_s$, $p_s$ and $T_s$ and advances to the next timestep if the check is passed, it stops the simulation otherwise. We now move to describe each of the individual steps in greater detail.

\subsection{The hydrodynamic substep}
\label{sec:numerical_hydro}

The hydrodynamics module solves the sub-system 
\begin{equation}
    \pderiv{\vec{U}_{s}}{t} + \nabla F_{s} = \vec{G}_{s}
\end{equation}
where
\begin{align}
    F_s &= \left\{ \rho_s u_s, ~ \rho_s u_s^2 + p_s, ~ u_s (E_s + p_s) \right\} \\
    \vec{G}_s &= \left\{0,~ -\rho_s \nabla \Phi, ~-\rho_s u_s \nabla \Phi_s \right\} 
\end{align}  
using the well-balanced Godunov method of \citet{Kappeli2016}. In this approach the conserved variables are updated according to
\begin{align}
    \vec{U}^{n+1}_{s,i} &= \vec{U}^{n}_{s,i} + \mathcal{L}_{\mathcal{H},i}(\vec{U}^n_s) \Delta t \label{eq:hydroupdate1} \\
    &=\vec{U}^{n}_{s,i} - \Delta t \frac{\left[A_{i+1/2} \vec{F}^n_{s,i+1/2} - A_{i-1/2} \vec{F}^n_{s,i-1/2}\right]}{V_i} + \vec{\tilde{G}}^n_{s,i}. \label{eq:hydroupdate2}
\end{align}
The interface fluxes, $\vec{F}^n_{s,i-1/2} \equiv \vec{F}(\vec{W}^n_{{\rm L},s,i-1/2},\vec{W}^n_{{\rm R},s,i-1/2})$ are obtained from the primitive variables ($\vec{W}_s = \{\rho_s, ~u_s,~ p_s\}$) evaluated at the left and right side of the interface using the HLLC Riemann Solver \citep{Toro1994}. This solver shows exceptional behaviour in the standard shock tube tests \citep{toro2009} and is particularly well-suited to be used in conjunction with the well-balancing as presented by \cite{Kappeli2016}, as it reproduces standing and slow-moving contact waves particularly well. For a uniform grid, \citet{Kappeli2016} propose that the reconstruction should satisfy the relation,
\begin{align}
    - \frac{\left[p^n_{s,i+1/2} - p^n_{s,i-1/2}\right]}{\Delta r} + \frac{\rho^n_{s,i} + \rho^n_{s,i+1}}{2}\frac{\Phi_{i+1} - \Phi_{i}}{\Delta r} = 0. \label{eq:hydrostatic_equation}
\end{align}
This can be achieved by extrapolating the cell centred pressures to the interfaces via 
\begin{align}
    p^n_{s,i}(x_{i\pm 1/2}) = p^n_{s,i} + \rho^n_{s,i} \frac{\phi_i - \phi_{i \pm 1}}{2},
\end{align}
and using $\vec{W}_{{\rm L},s,i-1/2} = \{\rho_{s,i-1}, u_{s,i-1},  p_{i-1}(x_{i-1/2})\}$ and $\vec{W}_{{\rm R},s,i-1/2} = \{\rho_{s,i}, u_{s,i},  p_{i}(x_{i-1/2})\}$. 

These expressions can be generalized to non-uniform grids and second-order spatial reconstructions as described in \citet{Kappeli2016} -- it is these generalizations, together with the Monotonized-Central slope limiter from \citet{Mignone2014}, that we use in the code.

The source term $\vec{\tilde{G}}_{s,i}$ includes the gravitational source term and an additional $-2p/r$ correction (in spherical symmetry) that appears when pulling $\partial P/\partial r$ into the differential operator in conservative form. The source term must be computed carefully to ensure that hydrostatic equilibrium is properly maintained, which we do via
\begin{equation}
    \vec{\tilde{G}}_{s,i} = -\begin{bmatrix}
            0 \\ 
            \rho_{s,i} \\
            u_{s,i} \rho_{s,i} \\
        \end{bmatrix}
        \left.\pderiv{\Phi}{r}\right|_i + a_{{\rm R}, i} 
        \begin{bmatrix}
            0 \\ 
            p^*_{i+1/2}\\
            0 \\
        \end{bmatrix}
        - a_{{\rm L},i}
        \begin{bmatrix}
            0 \\ 
            p^*_{i-1/2}\\
            0 \\
        \end{bmatrix},
\end{equation}
where $p^*_{i\pm1/2}$ is the pressure at the interface returned by the Riemann Solver and $\left.\pderiv{\Phi}{r}\right|_i$ is computed as in Appendix A of \citet{Kappeli2016}. The factors $a_{{\rm L/R},i}$ are essentially the difference between $\pderiv{r^2P}{r}/r^2$ and $\deriv{P}{r}$ , i.e.
\begin{align}
    a_{{\rm R}, i} &= \frac{A_{i+1/2}}{V_i} - \frac{1}{r_{i+1/2}-r_{i-1/2}} \\
    a_{{\rm L}, i} &= \frac{A_{i-1/2}}{V_i} - \frac{1}{r_{i+1/2}-r_{i-1/2}}.
\end{align}

To achieve second-order accuracy in time, we use the strong stability preserving (SSP)  Runge-Kutta time stepping scheme \citep{Gottlieb2001}:
\begin{align}
    \vec{U}^{*}_{s,i} &= \vec{U}^{n}_{s,i} + \mathcal{L}_{\mathcal{H},i}(\vec{U}^n_s) \Delta t, \\
    \vec{U}^{n+1}_{s,i} &= \vec{U}^{n}_{s,i} + \frac{1}{2}  \left[\mathcal{L}_{\mathcal{H},i}(\vec{U}^n_s)+\mathcal{L}_{\mathcal{H},i}(\vec{U}^*_s)\right] \Delta t. \label{eqn:step2}
\end{align}
Note that we also apply the friction and collisional heat exchange terms to $\vec{U}^*_{s,i}$ before evaluating the second step (\autoref{eqn:step2}).

\subsection{The friction and collisional heat exchange substeps}
\label{sec:numerical_friction}

In the friction step we solve the subsystem
\begin{align}
    \pderiv{\rho_s u_s}{t} &= \rho_s \sum_{s'} \alpha_{ss'} (u_{s'} - u_{s}). \\
    \pderiv{\rho_s e_s}{t} &= \rho_s \sum_{s'} \frac{\alpha_{ss'} m_{s'}}{m_s + m_{s'}} (u_{s'} - u_{s})^2 \label{eqn:drag_en}
\end{align}
Since $\alpha_{ss'} \propto \rho_{s'}'$ the drag term can be large in dense regions of the atmosphere and the friction step can be very stiff. Therefore we solve the for the new velocity using a semi-implicit method in which $\alpha_{ss'}$ is evaluated at $t_n$, whereas $u_s$ is evaluated at $t_{n+1}$
\begin{align}
    \frac{u_{s,i}^{n+1} - u_{s,i}^n}{\Delta t} = \sum_{s'} \alpha^n_{ss',i} ~ (u^{n+1}_{s',i} - u^{n+1}_{s,i}) \label{eqn:implicit_drag}
\end{align}
and $\rho_s$ is held constant. We assume here that $\alpha_{ss'}$ is constant over the timestep, an approximation which has been shown by \cite{benitez-llambay2019} to be acceptable. This is a linear system of equations coupling the velocities of all species in cell $i$, which we solve using the \textsc{eigen} linear algebra package. Similar approaches to treating drag forces were taken by \citet{Toth1995}, \citet{Stone1997}, and \citet{BL2019}.

Since energy is conserved under the action of drag forces, it is important to evaluate \autoref{eqn:drag_en} in such a way that the numerical scheme also conserves energy. To achieve this we use the fact that the change in kinetic energy of two species $s$ and $s'$ arising due to the interaction between these species should only heat those two species. Taking the force due to the interaction of species $s$ with $s'$ to be $\rho_s \alpha^n_{ss',i} (u^{n+1}_{s',i} - u^{n+1}_{s,i})$ for consistency with \autoref{eqn:implicit_drag}, the contribution to the change in kinetic energy of $s$ by this interaction during the timestep is 
\begin{equation}
W_{ss',i} = \Delta t \rho_{s,i} \alpha_{ss',i}(u^{n+1}_{s',i} - u^{n+1}_{s,i})\frac{(u^{n}_{s,i} + u^{n+1}_{s,i})}{2}
\end{equation}
and $W_{s's,i}$ follows by symmetry.

Since the change in $e_{s,i}$ and $e_{s',i}$ due to the drag force acting between them should equal $W_{ss',i} + W_{s's,i}$, this implies the following form for the energy update, 
\begin{align}
    (\rho e)_{s,i}^{n+1} =& (\rho e)_{s,i}^n  \nonumber \\&+ \Delta t \rho^n_{s,i} \sum_{s'} \frac{\alpha^n_{ss',i} m_{s'}}{m_s + m_{s'}} (u^{n+1}_{s',i} - u^{n+1}_{s,i})(u^{n+1/2}_{s',i} - u^{n+1/2}_{s,i}), \label{eqn:drag_en_num}
\end{align}
where $u^{n+1/2}_{s,i} = (u^{n}_{s,i} + u^{n+1}_{s,i})/2$. We then set $E^{n+1}_{s,i} = \frac{1}{2}\rho^n_{s,i} (u^{n+1}_{s,i})^2 + (\rho e)^{n+1}_{s,i}$. It is straightforward to verify that  \autoref{eqn:implicit_drag} and \autoref{eqn:drag_en_num} together conserve energy by comparing $\sum_s E^n_{s,i}$ and $\sum_s E^{n+1}_{s,i}$.

We treat the collisional heat exchange term,
\begin{equation}
\pderiv{E_s}{t} = \rho_s \sum_{s'} \frac{\alpha_{ss'}}{m_s + m_{s'}} 3 k_{\rm B} (T_{s'} - T_{s}),
\end{equation}
in an analogous way. By holding $\rho_s$, $\alpha_{ss'}$, and $u_s$ constant we can write
\begin{equation}
c_{v,s} \frac{T^{n+1}_{s,i} - T^{n}_{s,i}}{\Delta t}  = \sum_{s'} \frac{\alpha^n_{ss',i}}{m_s + m_{s'}} 3 k_{\rm B} (T^{n+1}_{s',i} - T^{n+1}_{s,i}), \label{eqn:dT_implicit}
\end{equation}
where we have used $e_{s,i} = c_{v,s} T_{s,i}$ such that $c_{v,s}$ is the heat capacity per unit mass of species $s$. In cases where we neglect radiation transport we again solve this linear equation using the \textsc{eigen} linear algebra package. When radiation transport is modelled we instead include this term in the radiation substep (\autoref{eq:radtrans_Ts_implicit}).

\subsection{The thermal radiation transport substep}
\label{sec:numerical_radiation}

\subsubsection{Opacities and solar absorption}

Before going into details of the radiation transport solver, we reiterate the definitions of three customarily used mean-opacities. Given a wavelength-resolved opacity function $\kappa_{\nu}(\rho,T)$, one can define the Rosseland-mean $\kappa_{R}$ as
\begin{align}
    \kappa_R^{-1}(\rho,T) =  \frac{\int_0^{\infty} d\nu \, \kappa_{\nu}^{-1}(\rho,T) \partial_T B_{\nu}(T) }{\int_0^{\infty} d\nu \partial_T B_{\nu}(T)} \label{eq:opacities_rosseland}
\end{align}
where $B_{\nu}(T)$ is the Planck-function as function of temperature. The single-temperature Planck-mean is 
\begin{align}
    \kappa_{\rm P}(\rho,T) =  \frac{\int_0^{\infty} d\nu \, \kappa_{\nu}(\rho,T)\, B_{\nu}(T) }{\int_0^{\infty} d\nu B_{\nu}(T)}  \label{eq:opacities_planck}
\end{align}
and the two-temperature Planck-mean, or solar opacity is 
\begin{align}
    \kappa_{\odot}(\rho,T,T_{\odot}) =  \frac{\int_0^{\infty} d\nu \, \kappa_{\nu}(\rho,T)\, B_{\nu}(T_{\odot}) }{\int_0^{\infty} B_{\nu}(T_{\odot})}  \label{eq:opacities_solar}
\end{align}
Note that with those definitions, it is $\kappa_R$ that determines whether a cell is optically thin or thick towards its own radiation, furthermore it is $\kappa_{\rm P}$ which locally couples the gas internal energy equation with the radiative energy equation, and $\kappa_{\odot}$  determines the solar absorption in the $\Delta S_s$-terms. Values for these three types of opacities can be found tabulated for gas-mixtures relevant to planet formation in \citet{freedman2014} and \citet{malygin2014} as functions of $\rho$, $T$ and $T_{\odot}$, but for the sake of simplicity we use $\kappa_{\rm P}=\kappa_R\neq\kappa_{\odot}$ not being functions $\rho$, $T$ and $T_{\odot}$ in this paper. We stress however that in future applications, with gases at arbitrary mixing ratios of species, it is necessary to know the per-species contributions to those opacities. 

As the attenuation of solar radiation per cell is nonlinear with their contributions w.r.t individual species' opacities, one has to take an indirect approach in order to compute the amount of stellar heating per species.
First, the solar heating function per species $S_s$, is computed from determining the total attenuation of a band flux in $b$ per cell $i$

\begin{align}
F^{b}_{\odot, i} = F^{b}_{\odot} \cdot exp(-\tau^b_{i}),    
\end{align}
with the total optical depth until the cell boundary
\begin{align}\tau^b_{\odot, i} = \int^{r_{i+1/2}}_{\infty} dr \sum_s^{\rm species}\rho_{\odot, s,i} \kappa^{b}_{\odot, s,i}
\end{align}
and between the cell walls
\begin{align}\Delta \tau^b_{\odot, i} = \Delta r \sum_s^{\rm species}\rho_{s,i} \kappa^{b}_{\odot, s,i}
\end{align}

The attenuation of radiation, i.e. the total photon energy deposited inside a cell is
\begin{align}
\Delta S^{b}_{\odot, i} = \frac{\partial F^{b}_{\odot, i} }{\partial r } (r_i) \approx F^{b}_{\odot} \times exp(-\tau^b_{\odot, i}) \, \frac{1-\exp(\Delta\tau^{b}_{\odot, i})}{\Delta r}.
\label{eq:dS_definition}
\end{align}
The latter is the cell-average of the local quantities in Eqn. \ref{eq:stellar_heating_1}, \citep[see also ][]{mellema2006} and the correct form for both non-ionising and ionising radiation. The complication for ionising radiation is that in between computing the opacities and the optical depths, one needs to find the density self-consistently with the ionisation rates.
In the cases for which we consider optically thick heating of the atmosphere from the planetary surface, we model a flux $\sigma T_{\rm int}^4$ by depositing the energy in the first active cell of size $\Delta r$, resulting in an additional contribution to the heating in this cell of 
\begin{align}
\Delta S^{b}_{\odot, 0} \mathrel{+}= \frac{\sigma T_{\rm int}^4}{\Delta r}.
\label{eq:dS_tinternal}
\end{align}
Next, the heating per species in this band and cell is reconstructed in a manner that conserves the photon flux, via its fractional contribution to the total optical depth for a given cell and band, as 
\begin{align}
\Delta S^{b}_{\odot, s,i} =  \Delta S^{b}_{\odot, i} \times \frac{\rho_{s,i} \kappa^{b}_{\odot, s,i}}{\sum_s \rho_{s,i} \kappa^{b}_{\odot, s,i}}.
\label{eq:solar2}
\end{align}
Finally, the total solar heating $S_{\odot, s,i}$ for a species in a given cell is computed as the sum over all its band heating contributions, i.e. 

\begin{align}
\Delta S_{\odot,s,i} = \sum_b^{\rm bands} \Delta S^{b}_{\odot, s,i}.
\label{eq:solar3}
\end{align}

The optical depths for ionising radiation are computed in the same way, but have to be solved self-consistently with the change in a species number density. Ionising heating terms are furthermore assigned to their product species, after being computed on the basis of their reactant species' number densities (e.g. the reaction $\rm H+h\nu\rightarrow H^{+}+e^{-}$, will have its heating rate proportional to $\kappa_{\rm H}\rho_{\rm H}$, but the resulting heating rate must be split among $\rm H^{+}$ and $\rm e^{-}$). This is described further in the section on ionising radiation.
Once the opacities and heating source terms are known, one can proceed to the solution of the energy equations.\\

\subsubsection{The thermal radiation solver}

The radiation transport substep updates the internal energy density and mean radiative energy simultaneously. As radiative evolution timescales can be shorter to any interesting hydrodynamic ones by many orders of magnitude, we use an implicit solution method. Both the $J^b$ and $T_s$ variables have to be evaluated at the advanced timestep $n+1$ for energy to be exactly conserved. In our implementation however, opacities and stellar source function are taken at the retarded timestep, $n$. In our test problems this seems to be sufficient for reasonable numerical stability, and in fact even the literature with non-constant opacities \citep{kuiper2010, vaytet2012, bitsch2013a} this does not seem to be a problem as long as the opacity is not a strong function of the temperature. Additionally we include the collisional heat-exchange terms in the internal energy update part of the radiation solver rather than solving for them separately via \autoref{eqn:dT_implicit}. 

The internal energy part of the total energy equation is thus
\begin{equation}
    \frac{\partial \rho_s  e_s}{\partial t} = \Gamma_s +  \rho_s \sum_{s'} \frac{\alpha_{ss'}}{m_s + m_{s'}} 3 k_{\rm B} (T_{s'} - T_{s}),
\end{equation}
which is solved simultaneously with Eqn. \autoref{eqn:radiation_equation} for the mean radiation intensity.
For the solution of the radiation transport problem, we use the temperature form of the internal energy equation, found via $e_s = c_{v,s} T_s$, which then is

\begin{align}
    c_{v,s}\frac{T^{n+1}_{s,i}-T^{n}_{s,i}}{\Delta t} =& 4\pi \sum_{b}^{\rm bands} \kappa^{b,n}_{\mathrm{P}, s,i} \left[ J^{b,n+1}_{i} - f^b(T^n_{s,i}) B(T^{n+1}_{s,i}) \right] \nonumber \\
    & \qquad +\frac{S^{n}_{\odot,s,i}}{\rho^n_{s,i}} + \sum_{s'} \frac{\alpha^n_{ss',i}}{m_s + m_{s'}} 3 k_{\rm B} (T^{n+1}_{s',i} - T^{n+1}_{s,i}) \label{eq:radtrans_Ts_implicit},
\end{align}

where the mean intensity relates to the energy density via, $4\pi J=c\, E_{\rm rad}$ and $B(T)=\sigma T^4/\pi$. To solve this equation we linearize the $B(T)$ term following the approach of \citet{commercon2011}, i.e. $(T^s_{i,m+1})^4 = 4 T^{n+1}_{s,i} (T^n_{s,i})^3 - 3(T^n_{s,i})^4$. The fraction $f^b(T_s)$ is tabulated at the beginning of each simulation using the Planck-integral and is fixed for the chosen band structure, from $10^{-10}$K to $10^{10}$K. Since $\sum_b f_b(T)=1$ for all temperatures, this guarantees energy conservation.

The discretized form of \autoref{eqn:radiation_equation} is

\begin{align}
    \frac{1}{c} &\frac{J^{b,n+1}_{i} - J^{b,n}_{i}}{\Delta t} + \frac{1}{4\upi}\frac{\left[A_{i+1/2} F^{b,n+1}_{i+1/2} - A_{i-1/2} F^{b,n+1}_{i-1/2}\right]}{V_i} \nonumber \\
    & \qquad= - \sum_{s}^{\rm species}  \rho^n_{s,i} \kappa^{b,n}_{\mathrm{P},s,i} \left[J^{b,n+1}_{i} - f^b(T_s) B(T^{b,n+1}_{s,i}) \right]
    \label{eq:radtrans_jb_implicit}
\end{align}
where the closure relation for relating fluxes and mean energy density is the discrete flux-limited diffusion relation 

\begin{align}
F^{b,n+1}_{i-1/2} = -\frac{4\upi\lambda(\overline{R}^{b,n}_{i-1/2}) }{\overline{\rho \kappa_{\mathrm{R}}}^{b,n}_{i-1/2}} \left(\frac{J^{b,n+1}_{i-1} - J^{b,n+1}_{i}}{ {r}_{i-1} - r_{i} }\right)    
\end{align}

where the mean Rosseland optical depths between two cells is taken as $\overline{\rho \kappa_{\mathrm{R}}}^{b,n}_{i-1/2} = ({\rho \kappa}^{b,n}_{{\rm R},i-1} + {\rho \kappa}^{b,n}_{{\rm R},i})/2$. The mean photon-mean free path parameter in the flux-limiter, $\overline{R}^{b,n}_{i-1/2}$, is defined by \autoref{eq:radtrans_r_parameter} and is approximated via
\begin{align}
   \overline{R}^{b,n}_{i-1/2}=\frac{\xi}{\overline{\rho \kappa_{\mathit{R}}}^{b,n}_{i-1/2}} \frac{J^{b,n}_{i-1} - J^{b,n}_{i}}{J^{b,n}_{i-1}}  .
   \label{eq:radtrans_r_parameter_numerical_stable}
\end{align}
which is sensible as long as the radiative flux vector is pointing away from the planet. The latter expression can be easily generalized via a standard upwinding method. Again, we refer to Sect. \ref{sec:results_rad} to a discussion of the factor $\xi$.
\autoref{eq:radtrans_jb_implicit} together with \autoref{eq:radtrans_Ts_implicit} form a block-tridiagonal system of linear equations (i.e. an $N_{\rm cells} \times N_{\rm cells}$ matrix of blocks with size $N_{\rm species}$ + $N_{\rm bands}$). We solve these directly using the \textsc{eigen} linear algebra package to obtain the new temperatures and mean intensities.

\subsection{The photoionization substep}
\label{sec:numerical_ion}

We have implemented a flexible model for the ionization of multiple species, and use the process $H + \gamma \rightarrow p^{+} + e^{-}$, as a simple test model for the source and sink terms, $Q_{\rm s}$ and $Q'_{\rm s}$. Explicitly we include photoionization, collisional ionization and recombination, along with the associated heating, cooling and momentum exchange between the species. The net production/loss of hydrogen is given by
\begin{equation}
Q_{\rm H} = m_{\rm H} \pderiv{n_{\rm H}}{t} = m_{\rm H}\left\{ \alpha_{\rm HI}(T_{\rm e})\, n_{\rm e} n_{\rm p} - \left[C_{\rm HI}(T_{\rm e})\, n_{\rm e} +  \Gamma_{\rm HI}\right] n_{\rm H} \right\} \label{eqn:ionization}
\end{equation}
where $n_{s}$ is the number density of species, $s$, $\alpha_{\rm HI}(T_{\rm e})$ is radiative recombination coefficient and $C_{\rm HI}(T_{\rm e})$ is the collisional ionization coefficient. The ionization rate is computed according to

\begin{align}
    \Gamma_{HI}= \frac{\partial F^{b}_{\odot, i} }{\partial r } (r_i) \approx F^{b}_{\odot} \cdot exp(-\tau^b_{\odot, i}) \, \frac{1-\exp(\Delta\tau^{b}_{\odot, i})}{\Delta r}.
\end{align}

We use the fits to the Case B recombination rates from \citet{hui1997}. This approximation corresponds to the 'on-the-spot'-approximation, for which ground-state recombination photons are locally re-absorbed and do not further change the ionisation rates \citep{mellema2006}. The source terms for protons and electrons are given by $\pderiv{n_{\rm p}}{t} = \pderiv{n_{\rm e}}{t} = -\pderiv{n_{\rm H}}{t}$. To solve the above equations, we follow the C2-ray approach \citep{mellema2006,friedrich2012}, in which $n_{\rm e}$ and $\Gamma_{\rm HI}$ in \autoref{eqn:ionization} are replaced by their averages over the time-step. This ensures that the number of photons absorbed in the cell for heating is identical to the number of atoms that are photoionized. The resulting equations are implicit in $n_{\rm e}$, and are solved iteratively using the Brent method \citep[e.g.][]{press2002}. 

Once the new densities have been determined, the next step is to compute the momentum and energy exchange due to photoionization. Applying momentum conservation to each of the processes individually implies that
\begin{align}
Q'_{\rm H} &=  \dot{n}_{\rm R} (m_{\rm e} v_{\rm e} + m_{\rm p} v_{\rm p}) - \dot{n}_{\rm I} m_{\rm H} v_{\rm H} \label{eq:ionization_momentum_correction_hydrogen} \\
Q'_{\rm p} &=  \dot{n}_{\rm I} \frac{m_{\rm H} v_{\rm H}}{ 1 + m_{\rm e}/m_{\rm p}}  - \dot{n}_{\rm R} m_{\rm p} v_{\rm p} \label{eq:ionization_momentum_correction_protons}\\
Q'_{\rm e} &=  \dot{n}_{\rm I} \frac{m_{\rm H} v_{\rm H}}{ 1 + m_{\rm p}/m_{\rm e}}  - \dot{n}_{\rm R} m_{\rm e} v_{\rm e},\label{eq:ionization_momentum_correction_electrons}
\end{align}
where $\dot{n}_{\rm R} = \alpha_{\rm HI}(T_{\rm e})\, n_{\rm e} n_{\rm p}$ and $\dot{n}_{\rm I} =\left[C_{\rm HI}(T_{\rm e})\, n_{\rm e} +  \Gamma_{\rm HI}\right] n_{\rm H}$. We update the resulting change in momentum implicitly, i.e. via 
\begin{equation}
m_{\rm H} \frac{n_{\rm H}^{n+1} v_{\rm H}^{n+1} - n_{\rm H}^{n} v_{\rm H}^{n}}{\Delta t} = \dot{n}_{\rm R} \left(m_{\rm e} v_{\rm e}^{n+1} + m_{\rm p} v_{\rm p}^{n+1}\right) - \dot{n}_{\rm I} m_{\rm H} v_{\rm H}^{n+1},
\end{equation}
where $\dot{n}_{\rm R}$ and $\dot{n}_{\rm I}$ are replaced with the time-averaged rates computed in the first step of the photoionization routine. Similarly $n_{\rm H}^{n}$ and $n_{\rm H}^{n+1}$ are the number density before and after the first step in the photoionization calculation. This equation can be solved along with the equivalent expressions for the other species to determine the new velocities, $v_s^{n+1}$.

Now that the new density and velocity of each species has been computed we can evaluate the total change in kinetic energy, $\Delta E_{\rm k}$. To ensure energy conservation we add this to the internal energy of the protons and electrons in the proportions $1/(1 + m_{\rm p}/m_{\rm e})$ and $1/(1 + m_{\rm e}/m_{\rm p})$ respectively.

Finally, the heating and cooling due to photoionization are given by
\begin{align}
\Lambda_{\rm H} &= \dot{n}_{\rm R} m_{\rm p} C_{\rm p}  T_{\rm p} - \dot{n}_{\rm I} m_{\rm H} C_{\rm H} T_{\rm H} \label{eq:ion_heating_h}\\
\Lambda_{\rm p} &= \dot{n}_{\rm I} \left(\frac{m_{\rm H} C_{\rm H}  T_{\rm H}}{1 + m_{\rm e}/m_{\rm p}} + \frac{E_{\rm HI}}{1 + m_{\rm p}/m_{\rm e}}\right) - \dot{n}_{\rm R} m_{\rm p} C_{\rm p}  T_{\rm p} \label{eq:ion_heating_p}\\
\Lambda_{\rm e} &= \dot{n}_{\rm I} \left(\frac{m_{\rm H} C_{\rm H}  T_{\rm H}}{1 + m_{\rm p}/m_{\rm e}} + \frac{E_{\rm HI}}{1 + m_{\rm e}/m_{\rm p}}\right) - \Lambda(T_e) \label{eq:ion_heating_e}
\end{align}
The terms related to ionization (proportional to $\dot{n}_{\rm I}$) can be derived by considering energy and momentum conservation during ionization, where $E_{\rm HI}$ is the average energy injected per ionization. This middle term is the high-energy equivalent of Eqn. \ref{eq:solar3}. To prevent double counting of photons, we keep track of the high-energy optical depth separately and substract it from the non-high energy photons, individually per band.
During recombination, the excess kinetic energy is lost to radiation. We have assumed this energy comes entirely from the electrons due to their higher velocities\footnote{Note that a more sophisticated calculation of recombination cooling can only slightly change the partitioning of the recombination cooling between protons and electrons because the final energy of the neutral atoms formed is determined by energy conservation.}.  It can be verified by summing the three terms that the net heating rate is just the energy injected per ionization minus the total cooling. Note that $\Lambda(T_e)$ is the total cooling rate, including recombination cooling and Lyman-alpha cooling (which is most important for planet photoevaporation, \citealt{MurrayClay2009}).

Since the heating and cooling rates can be numerically stiff when the ionization or recombination time-scales are short, we evaluate them semi-implicitly in the cases that thermal radiation transport is switched off in the code, by solving a modified version of \autoref{eqn:dT_implicit}, i.e.
\begin{equation}
c_{v,s} \frac{\rho_{s}^{n+1} T'_{s} - \rho_{s}^{n} T^{n}_{s}}{\Delta t}  = \rho^{n+1}_s \sum_{s'} \frac{\alpha^n_{ss'}}{m_s + m_{s'}} 3 k_{\rm B}  (T'_{s'} - T'_{s}) + \Lambda'_s,
\label{eq:ion_heating_total}
\end{equation}
where $\Lambda'_s$ is evaluated using $T'$. This intermediate value of $T'$ is then used to specify the heating and cooling rate used in the radiation transport substep.
With now conclude our technical description of the solution algorithms and turn to show test results.

\section{Results - Hydrostatics and hydrodynamics}
\label{sec:results_hydro}

\subsection{Hydrodynamic tests without gravity}
\label{sec:results_hydro_nograv}

We conducted basic hydrodynamic tests to make sure that the hydrodynamic solver works correctly. These include the seven standard shock-tube tests as given in \citet{toro2009}, which we tested using  a cartesian domain on $x\in[0,1]$ with a resolution of 100 cells. We confirmed that the HLLC solver resolves all waves correctly, with the code reproducing analytic solution results and converging as one would expect for the first and second order solvers as the resolution is increased. We also checked that for injected single sound-wavess, the solution converges appropriately to the chosen order as in \cite{Stone2008}; second order for the method described here. Since all those results are standard in the literature, we merely confirmed the codes correct functionality, but do not present them here. These tests are run regularly as part of our standard test package, included on the git repository.

\begin{figure*}
\centering
   \includegraphics[width=1.00\textwidth]{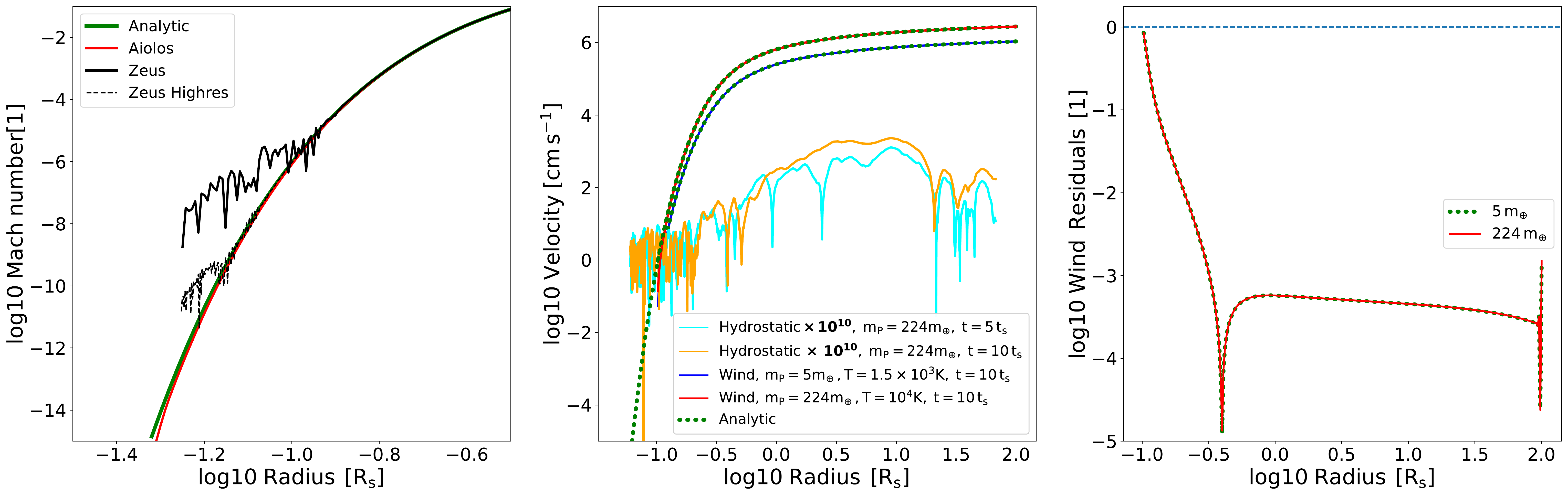}
\caption{Effects of the well-balancing for quasi-isothermal simulations. We compare our code and ZEUS \protect\citep{stonenorman1992} in their quality to balance a transonic, hydrodynamic wind solution in the deep, hydrostatic atmosphere (\textbf{Left}). The simulations are generated using an a Hot Jupiter-like exoplanet, an adiabatic, quasi-isothermal $\gamma_{\rm adi} = 1+10^{-8}$, the mean molecular mass $\rm m_{particle} = 2\,amu$,  numerical resolution of 100 cells per decade after $\rm 10 \,t_s$ (solid black, solid red line) and at double resolution after $\rm 10^3\,t_s$ (black dashed). Our well-balanced code compares well to the analytic solution by \protect\cite{cranmer2004} and only a systematic difference very deep in the atmosphere exists. This emphasizes the excellent behaviour of well-balanced codes in keeping hydrostatic balance deep in the atmosphere to more traditional approaches. 
Furthermore the wind solution is plotted over the entire simulation domain for the Hot Jupiter and a planet of drastically lower mass (\textbf{Middle}). The residuals w.r.t the analytic wind solutions (\textbf{Right}) for both planets are stable (identical for both masses) and better than $10^{-3}$ at the sonic radius. Deep in the atmosphere the residuals worsen as effect of adjusting to the wall boundary conditions we used.We also check the properties of the code to sustain an entirely hydrostatic solution over $\rm 10 \,t_s$ with open boundaries at large radii (\textbf{Middle}). The velocity solution in the fully hydrostatic case has been scaled up by $\times 10^{10}$ and shows only numerical noise in its velocity fluctuations, i.e. it remains hydrostatic grace to the well-balancing, and the amplitude of fluctuations only increases very slightly over time around the sonic point, but remains stable in the deep atmosphere. The wind and hydrostatic runs differ only in that the wind solution possesses a step function in the initial density profile at $r_s$.} 
\label{fig:wellbalancing}
\vspace*{+0.20cm}
\end{figure*}

\subsection{Hydrodynamics + gravity: well-balanced quasi-isothermal Parker winds}
\label{sec:results_hydro_grav}

In this subsection we report on the tests concerning the well-balancing of the scheme and its usefulness for hydrodynamic outflow simulations.

First, to highlight the necessity of schemes like presented in \cite{Kappeli2016}, we contrast our solutions in the deep hydrostatic atmosphere with those produced by a similar setup in the ZEUS code \citep{stonenorman1992} as in \cite{font2004}, only adapted for 1-D planetary atmospheres without the action of centrifugal and Coriolis terms. This highlights the well-behaved properties of the well-balancing scheme in the deep atmosphere, as can be seen seen in Fig. \ref{fig:wellbalancing}, left.

Second, we confirmed that a density profile of arbitrary entropy stratification, constructed by inverting Eqn. \ref{eq:hydrostatic_equation} for the density $\rm \rho^{n}_{s, i+1}(\rho^{n}_{s, i})$, remains static with the mach number $u/c_s \leq 10^{-8}$ over the entire simulation domain after 10 sound-crossing times in the simulation domain. The simulation domain was chosen to be between $r \in [0.1, 100] R_{s}$, where $R_{s}$ is the sonic radius. The boundary conditions employed are reflective at the inner radius and open boundaries with a hydrostatic pressure extrapolation at the outer boundary. Entropy stratifications tested were adiabatic (with $\gamma = 1.4$), quasi-isothermal (adiabatic with $\gamma = 1 + 10^{-8}$ for both masses of $5\, m_{\oplus}$ and $224\, m_{\oplus}$) and a `bumpy' temperature profile consisting of arbitrarily strong temperature inversion on top of the adiabatic temperature.

As shown in \cite{Kappeli2016}, the hydrostatic profiles should all remain stable if they are constructed to obey the discrete hydrostatic equilibrium, which is indeed the case. We show the evolution of the velocity fluctuations in Fig. \ref{fig:wellbalancing} (\textbf{Middle}) for the quasi-isothermal gas giant atmosphere. 
Deep in the atmosphere, at $r/R_s<0.1$ the hydrostatic equilibrium is kept extremely well, as can be seen in the small magnitude of the velocity in those profiles, with larger fluctuations at and above the sonic point. There is some nonzero growth of the velocity noise in this region of the high atmosphere. Considering that this is however the region that is dynamically most active in realistic settings, this slight growth in velocity fluctuations should never play any significant role for applied simulation cases.

We next compare the isothermal wind solutions provided by the simulation with those computed by \citet{parker1965}, with the formalism from \citet{cranmer2004} for a $5 m_{\oplus}$ and a $224 m_{\oplus}$ planet. The numerical wind solutions match the analytic solutions near perfectly (see Fig. \ref{fig:wellbalancing}, left) and that remains true for the low-mass planet when inspecting the relative and absolute residuals (see Fig. \ref{fig:wellbalancing}, right).

\section{Results - Friction}
\label{sec:results_drag}

For a purely isothermal gas mixutre, the coupling between multiple species is entirely given by their momentum exchange due to drag forces. Hence, to test the drag module included in our code, outlined in previous sections, we use a simple multiple-species-in-a-box approach, without the action of gravity, identical to the tests outlined in \cite{benitez-llambay2019}. A further test, including the action of gravity, serves to test our numerical results against the subsonic two-species drag approximation of \cite{zahnlekasting1986}.

\subsection{Drag without gravity}
\label{sec:results_drag_nodrag}

Here we conducted tests to confirm that the evolution of multiple fluids uniform density, velocity and pressure evolves as expected. The tests were identical to the one presented in \cite{benitez-llambay2019} and used 2,3 or 6 species. We confirmed that our algorithm agrees with the analytic solutions for constant friction coefficients, $\rm \alpha_{ij}$. Energy and momentum are conserved to better than one part in $10^{10}$ after a frictional time. We also the attainment of the correct steady-state mean velocity $\bar{v}$
\begin{align}
\bar{v} = \frac{\sum_s \rho_s v_s(t=0)}{\sum_s \rho_s}.    
\end{align}

\begin{figure*}[h]
\hspace*{-0.5cm}        
  \centering
   \includegraphics[width=1.0\textwidth]{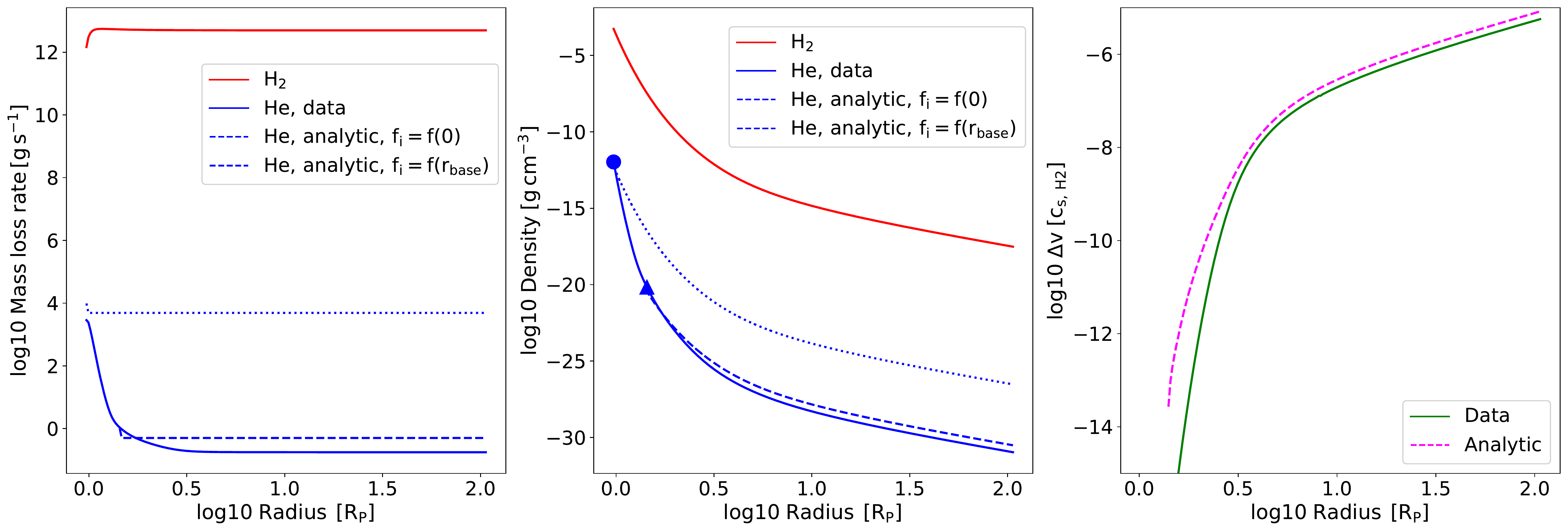}
\caption{H/He coupled together via physical, i.e. non-constant collision constants, with He being a minor species. We compare mass-loss rates, density profiles and velocity difference with the analytic solutions, based on two different choices of the lower radius $r_0$ in Eqn. \ref{eq:zk_analytic}. This highlights that when comparing a dynamic simulation with a static solution, one has to choose $r_0$ such that the lower radius $r_{\rm base}$ corresponds to the flow-time during the simulation time. The velocity differences between the two species are discussed in the text.} 
\label{fig:drag_domain}
\vspace*{+0.20cm}
\end{figure*}

\begin{figure*}[h]
\hspace*{-0.5cm}        
  \centering
   \includegraphics[width=1.0\textwidth]{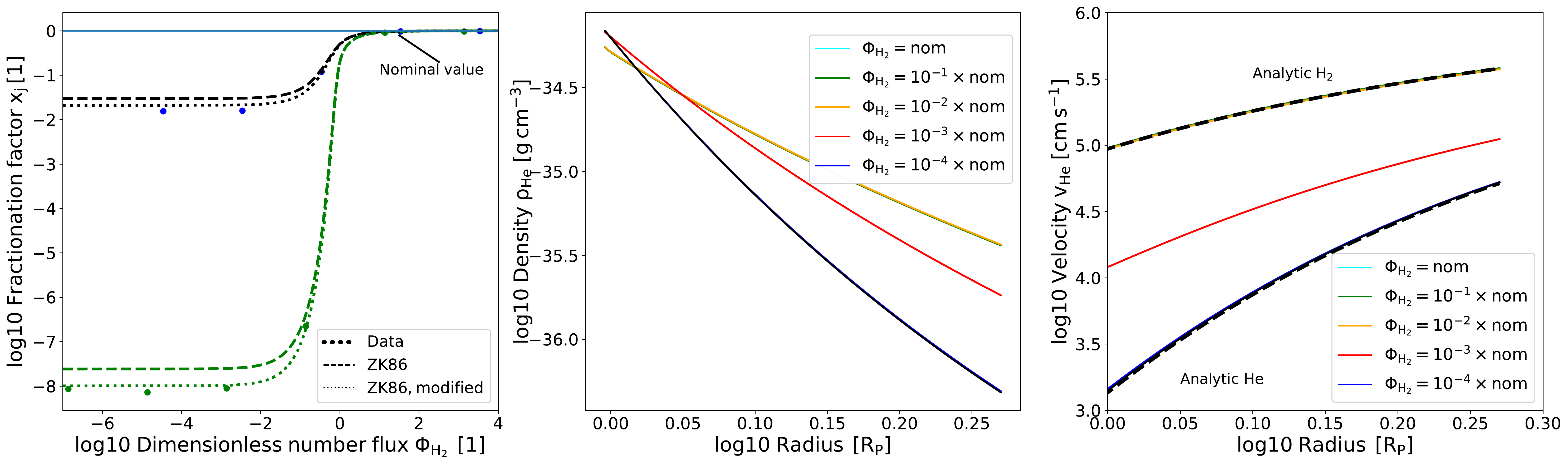}
   
\caption{Same as above, but at a planet with $m=224m_{\oplus}$ at $T=8000$K, varying the hydrogen density, i.e. the hydrogen escape flux with the intention of showcasing the transition from well-coupled to decoupled two-species solutions for Helium (\textbf{Left}, black dashed) and a $m_j=10 \rm u$ species (\textbf{Left}, green dashed). Density profile and the velocity profile of He (\textbf{Middle} and \textbf{Right}) adjust from high hydrogen flux to low hydrogen flux cases. It becomes evident that in the decoupled cases each species follows their own Parker-wind solution (\textbf{Right}).} 
\label{fig:drag_transition}

\vspace*{+0.20cm}
\end{figure*}

\subsection{Drag including gravity - Fractionation}
\label{sec:results_drag_grav}

The effects of inclusion of gravity on a single species have already been discussed above. With the addition of more species, e.g. a trace species, the dominant force contribution on those species can shift from drag-dominated to gravity-dominated anywhere in the simulation domain. We show that in both those limits our code produces correct results, using two species with masses $m_i$, $m_j$ and $m_j>m_i$.
In this case, \cite{zahnlekasting1986} (hereafter ZK86) have given approximate analytic results for the mass flux ratio at infinity. Furthermore, both in their work and here, non-constant collision coefficients of $\alpha_{ij} = k_B T n_j / (m_i b_{ij}) $ and the experimentally measured $b_{ij} = 5\times 10^{17} \times \left(T/300 K \right)^{0.75}$ are employed.
The two-species fractionation factor $x_j = \frac{n_i v_i}{n_j v_j} (r_s) \frac{n_j}{n_i} (0)$ i.e. the mass flux ratio at infinity, normalized to the density ratio at some base radius $r_0$, in this case is expressed as 

\begin{align}
    x_j = \frac{\mu \Phi_i - (\mu-1)}{\mu \Phi_i - (\mu-1)\exp{\left[ - \lambda_0  (\mu \Phi_i - (\mu-1) ) \right]}}
\label{eq:zk_analytic}
\end{align}
where
\begin{align}
    \mu = m_j/m_i \\
    \Phi_i = (n_i r^2) |_{r=r_{max}} \frac{k_B T}{Gm_p m_j b_{ij}} \\
    \lambda_0 = \bar{r}_s/r_0 \\
    \bar{r}_s = \frac{G m_p}{k_B T} \frac{m_i m_j}{m_i+m_j}.
\end{align}
We first compare the behaviour of the dragged solution for one simulation over the entire simulation domain. The results of this can be seen in Fig. \ref{fig:drag_domain}. The left and middle panel shows the behaviour of the density and the mass-flux $\rm \rho v r^2$ for primary ($\rm H_2$) and a secondary species ($\rm He$). Those panels showcase the effect of chosing the right base radius for the comparison of numerical with the analytic solution, e.g. Eqn. \ref{eq:zk_analytic}. When the time-dependent problem has not run into steady-state yet, chosing $r_0 = r_{min}$ can result in a faulty comparison. If one however computes $r_0$ such that it corresponds to the flow time of the parcel over the simulated time, denoted as $r_0 = r_{base}$, one obtains much better agreement in mass-flux, density profile, and velocity differences.

The velocity differences can be expressed in a generalization of Eqn. 9 in \cite{zahnlekasting1986} which gives, 

\begin{align}
    u_1-u_2 = \frac{k_B T n_2}{m_1  n b_{ij}} \left[
    \frac{n}{n_1 n_2}  \frac{\partial n_1}{\partial r}  \left(1-\frac{n_1}{n}\right) 
    - \frac{1}{n_2}     \frac{\partial n_2}{\partial r} \right. \\
    \left. +\frac{m_2-m_1}{m_1+m_2} \frac{1}{p_1} \frac{\partial p_1}{\partial r}
    +\frac{g (m_2-m_1)}{k_B T} \right]
    \label{eq:velocity_differences}
\end{align}
where $n=n_1+n_2$ and $p_1$ is the pressure of the mass-dominating species. The velocity differences from the simulation are then compared to the velocity difference that should arise from the simulations structure data plugged into Eqn. \ref{eq:velocity_differences}. They achieve acceptable agreement, given that Eqn. \ref{eq:velocity_differences} is derived in our, as well as in the work by \cite{zahnlekasting1986} via the hydrostatic approximation. We show in the next paragraphs that the approximation in \cite{zahnlekasting1986} needs a slight correction.

Next, the behaviour of the fractionation factor $x_i$ over a wide range of coupling parameters $\Phi_{H_2}$ is shown in Fig. \ref{fig:drag_transition}. Those simulations have been performed with open inner boundary conditions, which allows to relax the simulation to steady-state as the flow timescale throughout the domain is lower. We compare solutions for two different secondary species, He with $m_j=4\rm u$ and an unnamed other species with $m_j = 10\rm u$. In the left panel in Fig. \ref{fig:drag_transition} the transition of the secondary species escaping flux at infinity from well-coupled to uncoupled becomes quite obvious. 
The secondary species radial profiles change accordingly - at high $\Phi_{H_2}$, i.e. in the well-coupled regime, the velocity profile as well as the density scale height attained is that of the primary species, whereas in the uncoupled regime, the secondary species behaves independently. We emphasize that in the well-coupled regime the secondary escape flux is not identical to the primary escape flux, it merely becomes identical to the base density ratio $n_j/n_i (r=r_p)$.

We note further that in the uncoupled regime, Eqn. \ref{eq:zk_analytic} produces a slightly too high prediction for the secondary escape rate. The reason for this lies in the assumption by ZK86 that the secondary species remain subsonic, which increases its density at its sonic point.

We correct this by taking into account the note by ZK86 that in the correct limit, when both species experience supersonic escape, $u_j/u_i \rightarrow \sqrt{1/\mu}$ has to hold. Furthermore, in this limit, the species' solutions are separate Parker winds. Therefore, we modify Eqn. \ref{eq:zk_analytic} by adding a factor $\sqrt{\mu}$ into the denominator, resulting in
\begin{align}
    x_j = \frac{\mu \Phi_i - (\mu-1)}{\mu \Phi_i - \sqrt{\mu}(\mu-1)\exp{\left( - \lambda_0  (\mu \Phi_i - (\mu-1) ) \right)}},
\label{eq:zk_analytic_modified}
\end{align}
this is shown in Fig. \ref{fig:drag_transition} to be in much better agreement with our numerical findings. We take the good agreement between this modified formula and our numerical results as a sign of both being the correct, transonic solution. While analytically true, this uncoupled limit is physically suspicious, given that the two (or more) fluids are not collisionally  coupled to each other, but retain their transonic outflows. Under such conditions the individual fluids would likely not remain collision and the fluid approximation would break down, with the true escape rates being given by Jeans-like escape of particles \citep{volkov2016}.

\section{Results - Radiation transport}
\label{sec:results_rad}

Our radiation transport scheme is a multi-band, multi-species extension of the schemes presented in \cite{commercon2011} and \cite{bitsch2013a, lega2014}, inspired by the formalism presented in \cite{vaytet2012}. The scheme is implicit and by construction conserves the sum of radiative and internal energies over any time step.

Before presenting numerical results, we introduce a generalization of the popular model for irradiated, plane-parallel planetary atmospheres by \citet{guillot2010}. This  model is based on the idea that the radiation can be split into the incoming stellar irradiation and the re-radiated bands, which is the approach taken in \textsc{aiolos}. Here we have generalized the solution to include multiple bands for the stellar irradiation; the temperature as a function of optical depth is given by
\begin{align}
T^4 =& \, T_{\rm int}^4 \left(\frac{3 \tau}{4} + \frac{1}{4 f_{\rm H}}\right) +  \nonumber \\ &\sum_b \frac{3 \mu}{4} \, \frac{f S_{\odot,b}}{\sigma_{\rm SB}} \left[\frac{1}{3f_{\rm H}} + \frac{\mu}{\gamma_b} + \left(\frac{\gamma_b}{3\mu} - \frac{\mu}{\gamma_b}\right)\exp\left(-\tau \frac{\gamma_b}{\mu}\right)\right]. \label{eq:guillot_multi}
\end{align}
Here $T_{\rm int}$ is the internal luminosity of the planet expressed as a temperature, implemented via Eqn. \ref{eq:dS_tinternal}.

The irradiation temperature, a measure of a local radiation fields' energy content, is for the solar irradiation $T_{{\rm irr},b}^4= S_{\odot,b}/\sigma_{\rm SB}$, and the planetary equilibrium temperature for a zero albedo planet is related to $T_{\rm irr}$ via $T^4_{\rm eq} = T^4_{\rm irr}/4$. The optical depth in the thermal radiation band is $\tau$, and $\mu$ is the angle of incident radiation, assumed to be $1$ in \textsc{aiolos}. The parameters $\gamma_b = \kappa_{\odot, b}/\kappa_{\rm R}$ are the ratio of the opacity to stellar irradiation in the different bands to the opacity to the outgoing thermal radiation, which is assumed to be constant over radius and time. The factor of $f$ is included because \textsc{aiolos} is a 1-D code and represents the redistribution geometry of stellar irradiation over the full planet (see Appendix \ref{app:multi_guillot}). The parameter $f_{\rm H}$ is a boundary condition that relates the radiation flux to the radiation intensity at the top of the atmosphere, i.e., $f_{\rm H} = H/J$ (where $F= 4\upi H$). \citet{guillot2010} recommends $f_{\rm H}=1/2$ to match more detailed solutions for plane-parallel emission. However, the FLD approximation traditionally assumes $H=J$ in the optically thin limit, which is appropriate for regions far from the emission surface where the geometry is no longer plane-parallel. We have included a factor $\xi$ that corresponds to $1/f_{\rm H}$ in \textsc{aiolos}' flux-limiter (see Eqn. \ref{eq:radtrans_r_parameter}) and will thus compare our numerical solutions with both $\xi=1$ and $\xi=2$ to \citet{guillot2010}'s model.


A property of Eqn. \ref{eq:guillot_multi} is that a characteristic temperature in the profile is
 $T^4_{\rm eff} =  \left( T^4_{\rm int} + \frac{1}{4} T^4_{\rm irr} \right) = \left( T^4_{\rm int} + T^4_{\rm eq} \right)$, when choosing their parameter values as $\mu_{*}=1$(irradiation with zero angle w.r.t vertical), $f_H=1/2$ (outer boundary condition), $f_K = 1/3$ (Eddington factor) and $f=1/4$ (global average). 

Finally, the minimum temperature appearing in a temperature profile, $\rm T^4_{\rm min}$ lies between $T^4_{\rm eff}$ and $T^4_{\rm eff}/2$ \citep{parmentier2014} depending on the value of $\gamma$ (the latter for $\gamma \rightarrow \infty$, $\gamma \rightarrow0$ and the former for $\gamma \rightarrow2$). We discuss the conditions under which our numerical model can reproduce those low temperatures at intermediate altitudes in the next section, as this minimum temperature in the profile can form a cold trap \citep[][Chapt. 5]{catlingbook2017} and thus severely limit the escape rates (ref). The temperature in the upper atmosphere, in the limit $T(\tau\rightarrow 0)$ is termed  the skin temperature and is $T^4_{\rm skin} =\frac{1}{2} T^4_{\rm eff} + \frac{\gamma}{4} T^4_{\rm eq} $.


It is a general feature of double-grey models that the temperature minimum is $T_{\rm min}\leq T_{\rm eff}$ in the temperature profile $T(r)$, which remains true in dedicated, line-by-line high-resolution studies of atmospheric temperature profiles \citep[Fig. 18]{parmentier2015}.

\begin{figure}
\hspace*{-0.5cm}
\begin{subfigure}{0.50\textwidth} 
   \centering
   \includegraphics[width=1.0\textwidth]{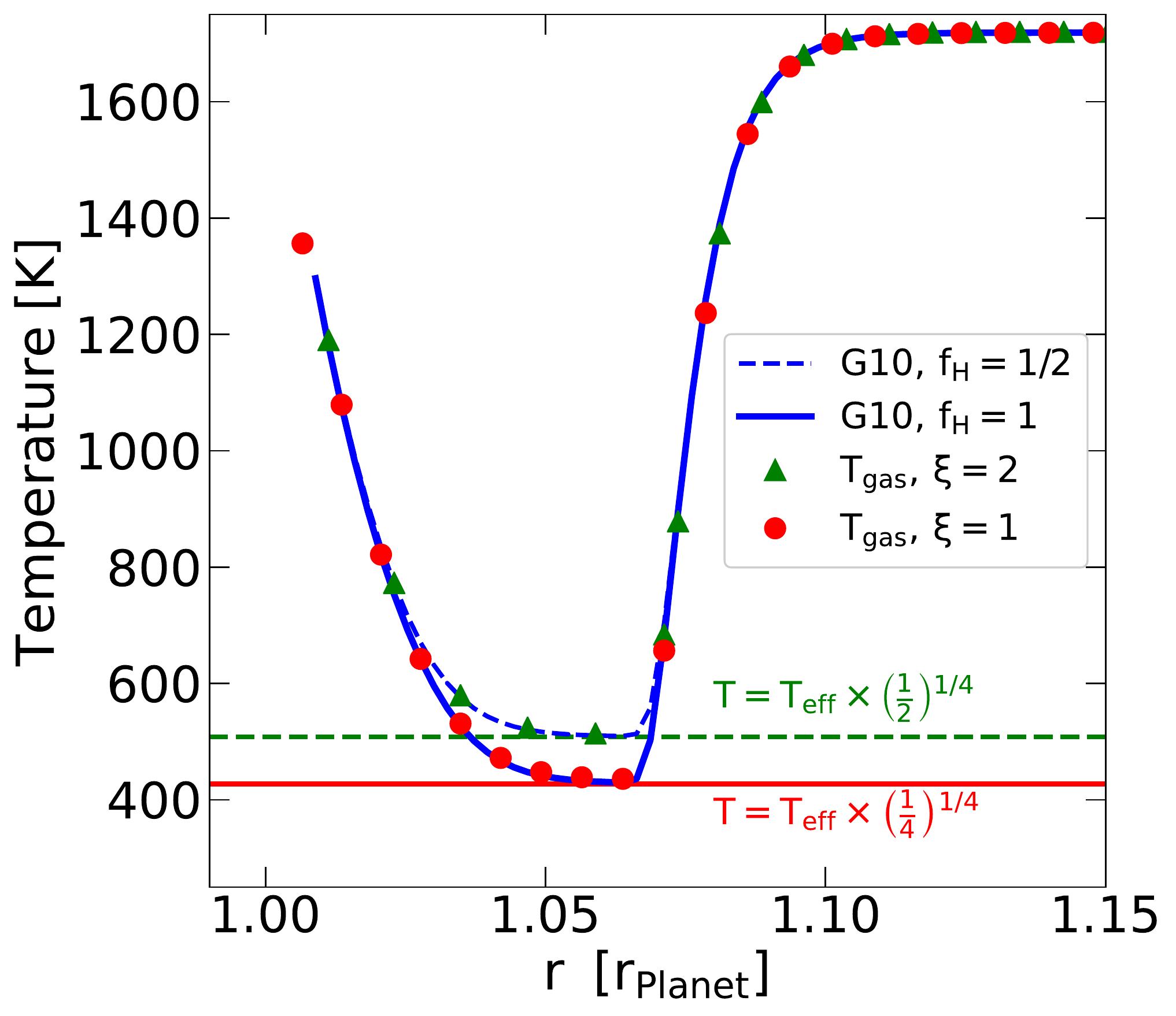}
\end{subfigure} 

\caption{Middle-atmospheric temperature profiles after $10^{13}$s in steady-state. We compare two solutions from \citep{guillot2010} varying $f_H = H/K$ and equivalently, $\xi$ in our simulations, with $\gamma=10^2$. It is clear that one needs to set $\xi=2$ in order to get the correct minimum temperature in the atmosphere, which is $T_{\rm eff}/2^{1/4}$ \citep{dobbs-dixon2012, parmentier2014}.} 
\label{fig:temperatures_analytic0}
    
\vspace*{+0.20cm}
\end{figure}

\subsection{A comment on the FLD factor $\xi$}
As noted previously, the factor $\xi$ that appears in our flux limiter (Eqn. \ref{eq:radtrans_r_parameter} and Eqn. \ref{eq:radtrans_r_parameter_numerical_stable}) determines the FLD flux in the optically thin limit. The nominal value of $\xi=1$ 
is widely used as standard in the literature and corresponds to a collimated beam of radiation, but it fails to reproduce the analytical solution for irradiated planetary atmospheres with the parameters $\mu_*, f_{\rm H}, f_{\rm K}, f$ as introduced before, see Fig. \ref{fig:temperatures_analytic0}. To obtain a match the FLD results and the analytic results we have to assume either $f_{\rm H}=1$ in the \cite{guillot2010}-solution (corresponding to collimated radiation in the optically thin regions) or rescale our mean-free path parameter in the FLD solution via the choice $\xi=2$. With $\xi=2$ our hybrid FLD code obtains the correct temperature minimum, $T_{\rm eff}\times\left(\frac{1}{2}\right)^{1/4}$, of the semi-grey plane-parallel model and matches the analytic temperature profile Fig. \ref{fig:temperatures_analytic0}. The re-scaling via $\xi=2$ simply accounts for the fact that, in a plane-parallel geometry, the radiation field is not collimated. Instead there is a significant contribution from rays travelling laterally through the atmosphere that means the flux is only $1/2$  of its free-streaming value \citep[e.g.][]{dobbs-dixon2012}. This is the main reason for the introduction of the $\xi$-parameter in our code, and the user can choose between those values, depending on the application. The comparison is shown in Fig. \ref{fig:temperatures_analytic0}.

It is important to note that $\xi = 2$ may not always represent the best choice. The results of plane-parallel fully line-by-line radiative transfer models show that non-grey effects can cause the minimum temperature to be below that of the \citet{guillot2010}-type models \citep{parmentier2015}. Although the non-grey effects could be accounted for by using a sufficient number of bands in the FLD calculation, one might find sufficient accuracy with fewer bands by tuning the choice of $\xi$. Another point of consideration is that the choice $\xi=2$ is appropriate for a plane parallel geometry -- far away from the planet the thermal radiation field must eventually become collimated, resulting in $\xi = 1$. The impact of this error will depend on $\gamma$: for large $\gamma$ the skin temperature is not much affected by the thermal radiation emitted by the planet since $T_{\rm skin}^4 \approx \gamma T_{\rm eq}^4/4$, but for small $\gamma$ the temperature far from the planet will be overestimated by a factor $\xi^{1/4}$ (1.189 for $\xi=2$).

\begin{figure*}
\hspace*{-0.5cm}
 \begin{subfigure}{0.32\textwidth} 
   \centering
   \includegraphics[width=1.0\textwidth]{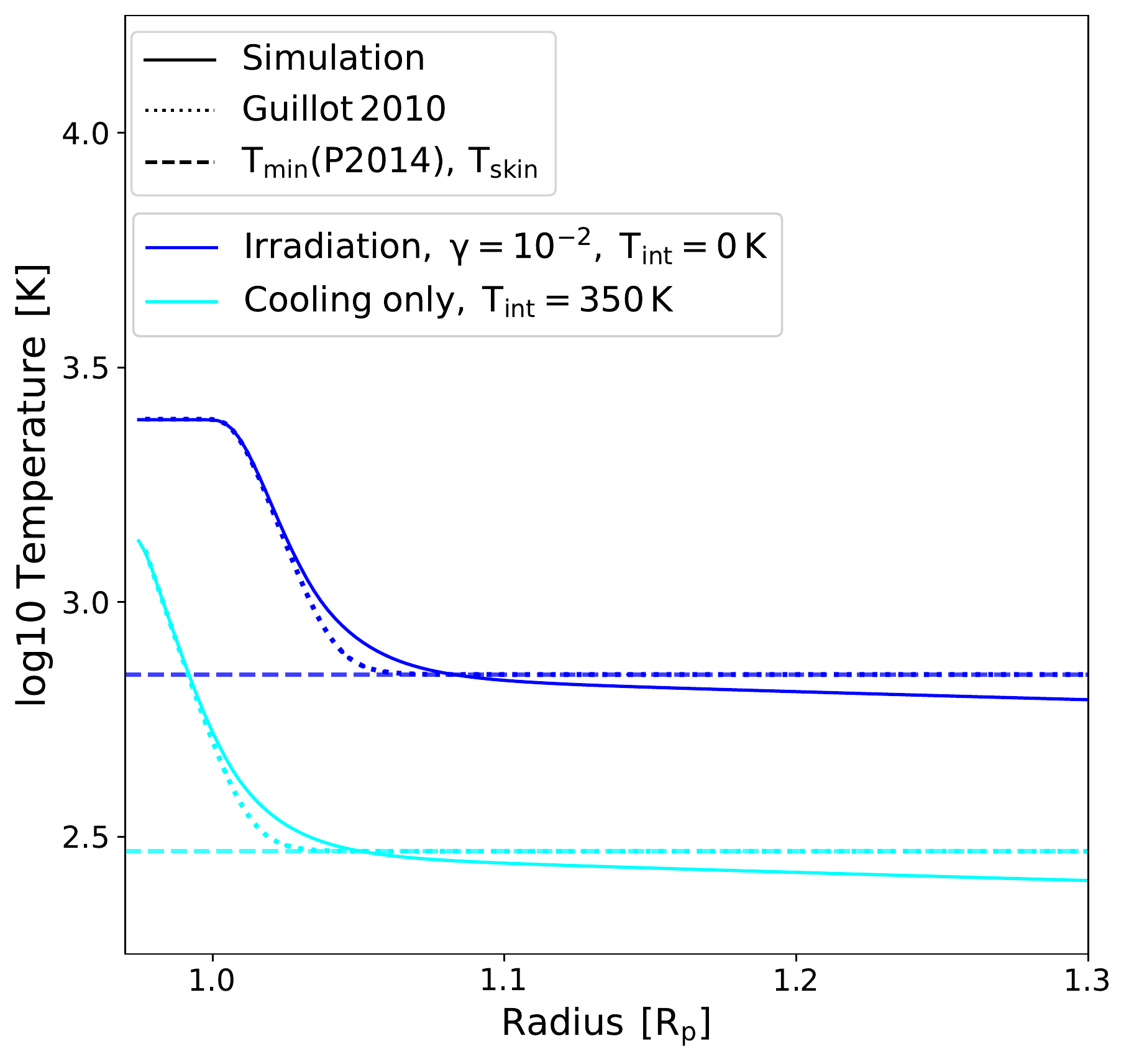}
\end{subfigure}%
\begin{subfigure}{0.32\textwidth} 
   \centering
   \includegraphics[width=1.0\textwidth]{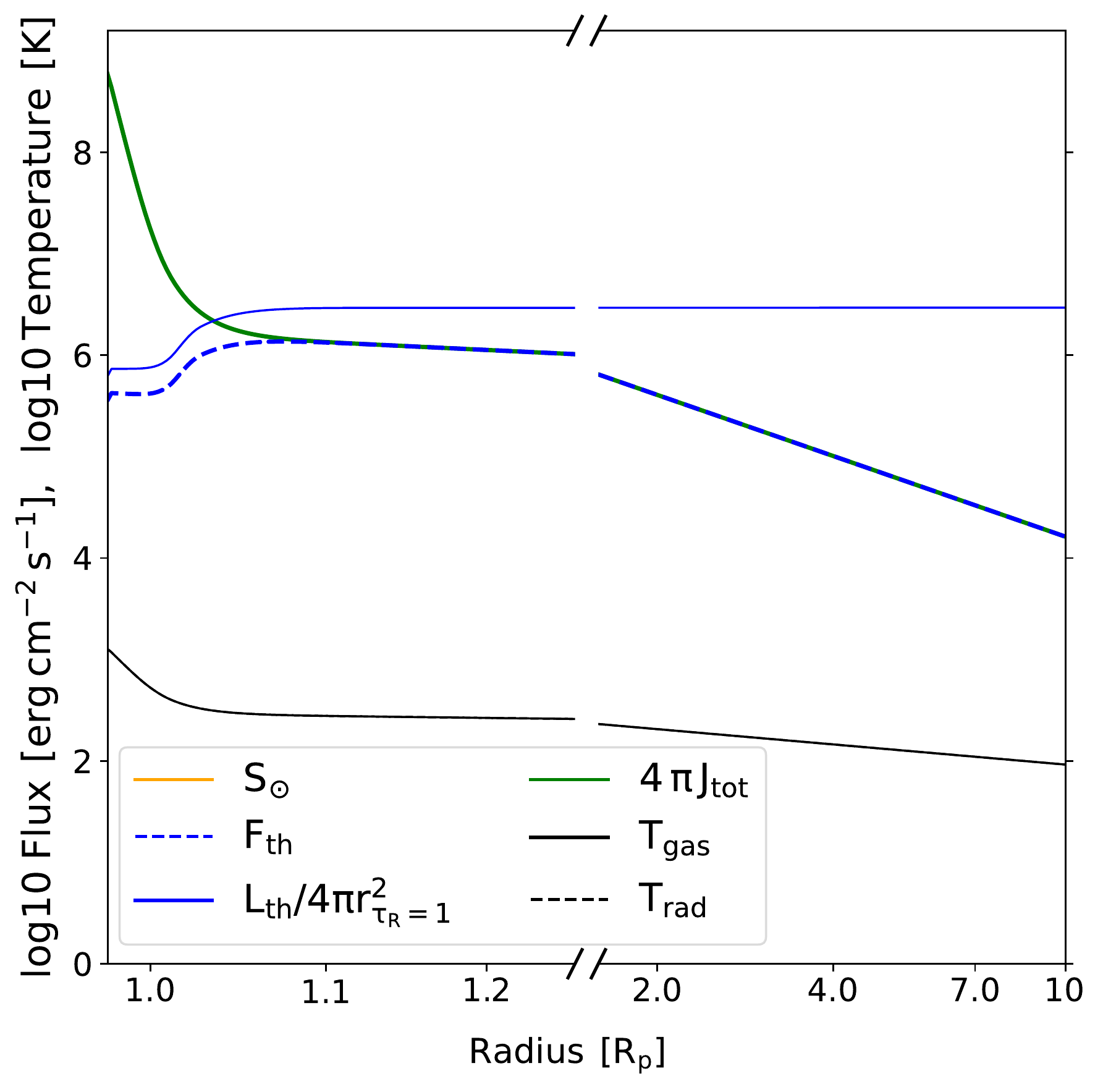}
\end{subfigure} 
\begin{subfigure}{0.32\textwidth} 
   \centering
   \includegraphics[width=1.0\textwidth]{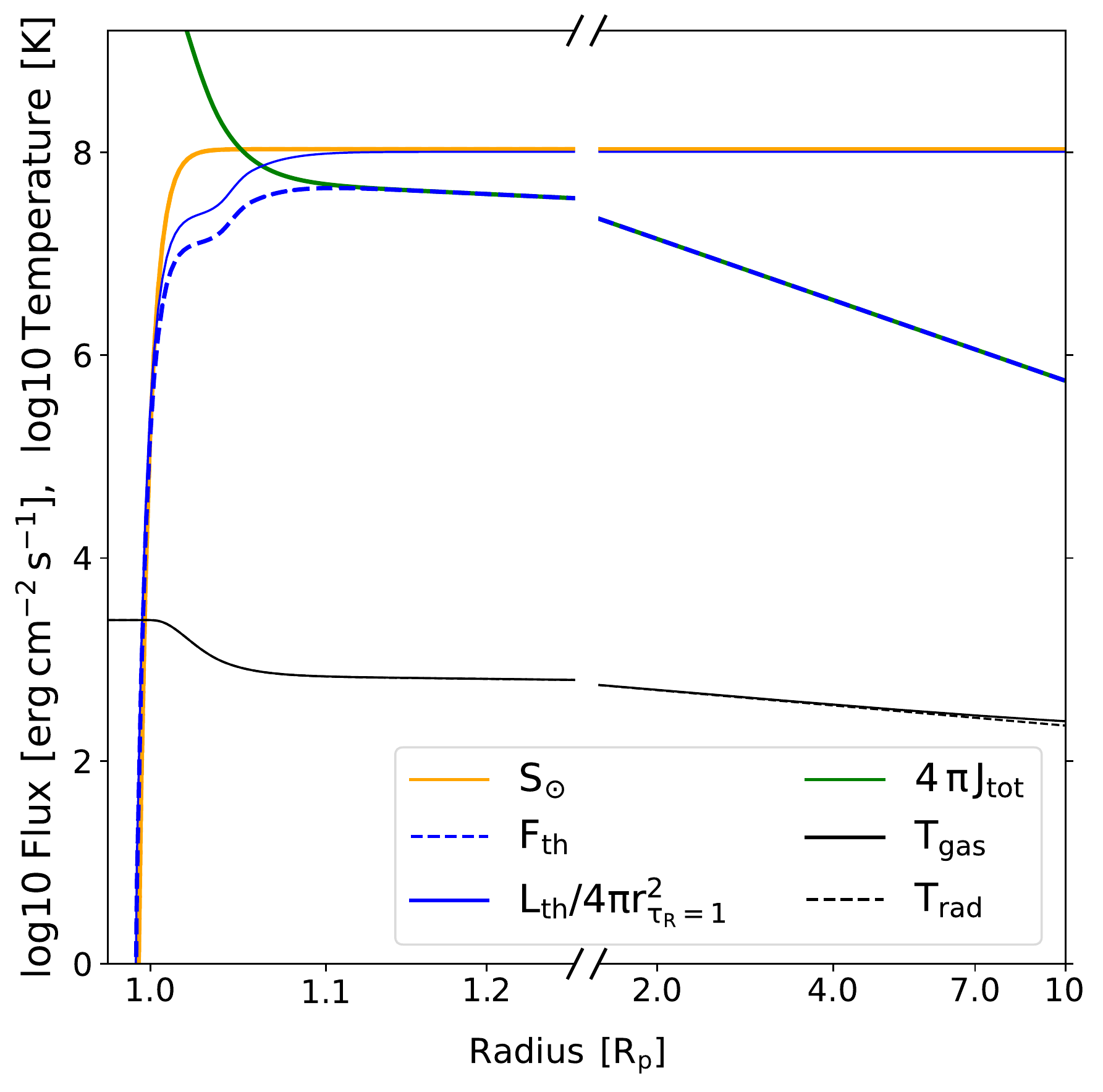}
\end{subfigure}
\caption{Comparisons for radiative quantities in steady-state after $3\times10^{13}s$. An atmosphere with only planetary luminosity (Cyan, \textbf{left} and \textbf{middle}) 
and $\gamma<1$ irradiation without planetary luminosity (Blue, \textbf{left} and \textbf{right}). The left panel compares the temperature profiles with those from \protect\cite{guillot2010} 
and shows the deviation from the \protect\cite{parmentier2014}-minimum temperature (see text).
Middle and right panels show equilibrium between ingoing and outgoing fluxes where $S_{\rm 0}=L_{\rm th}/4\pi r^2_{\tau=1}$, radiative equilibrium as $T_{\rm rad}=T_{\rm gas}$, 
the correct attainment of the optically thin limit as 
$F_{\rm th}=4\pi J_{\rm tot}$ and $F_{\rm th}\propto r^{-2}$.  } 
\label{fig:temperatures_analytic1}

\vspace*{+0.20cm}
\end{figure*}

\begin{figure*}
\hspace*{-0.5cm}
 \begin{subfigure}{0.32\textwidth} 
   \centering
   \includegraphics[width=1.0\textwidth]{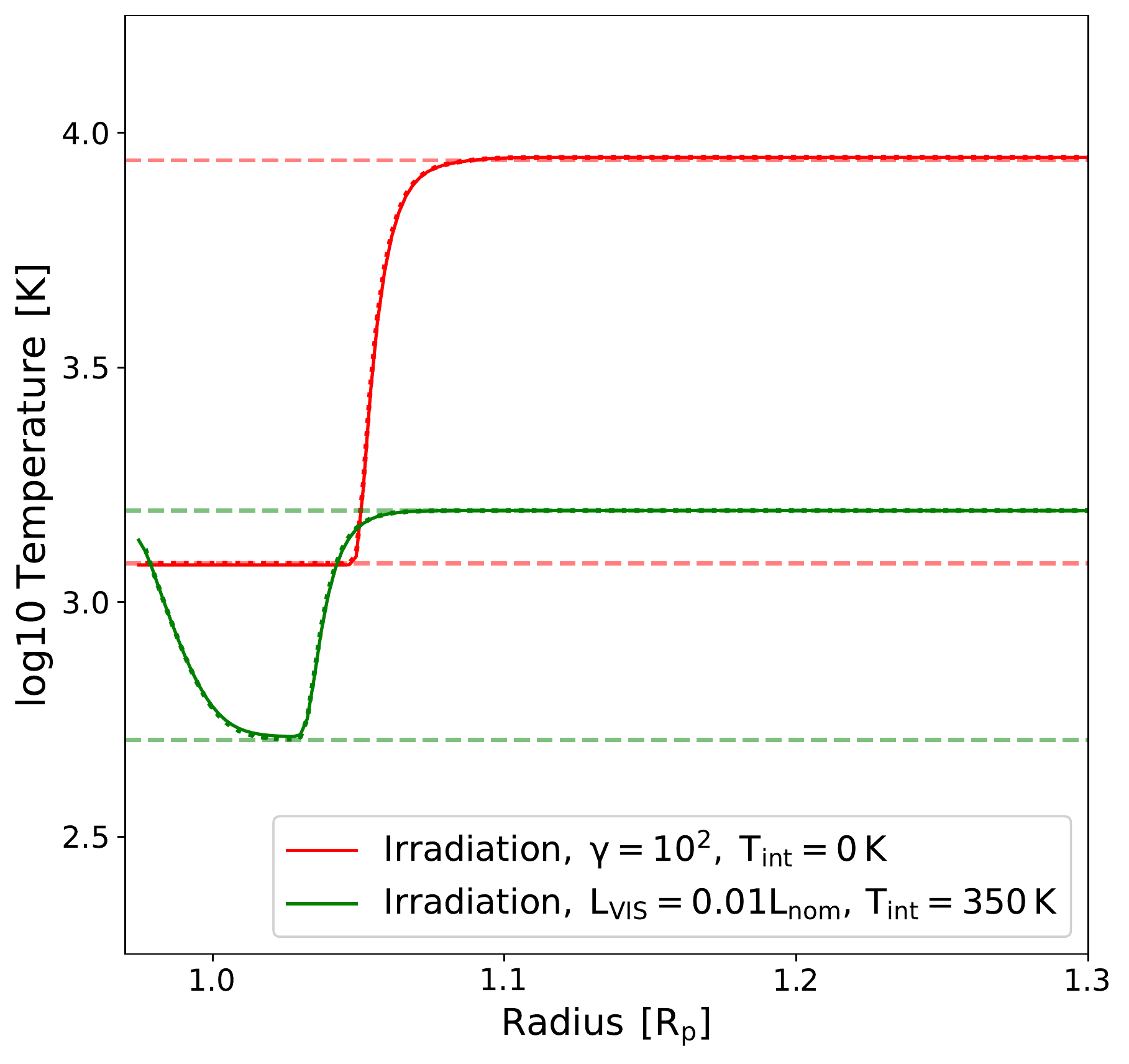}
\end{subfigure}%
\begin{subfigure}{0.32\textwidth} 
   \centering
   \includegraphics[width=1.0\textwidth]{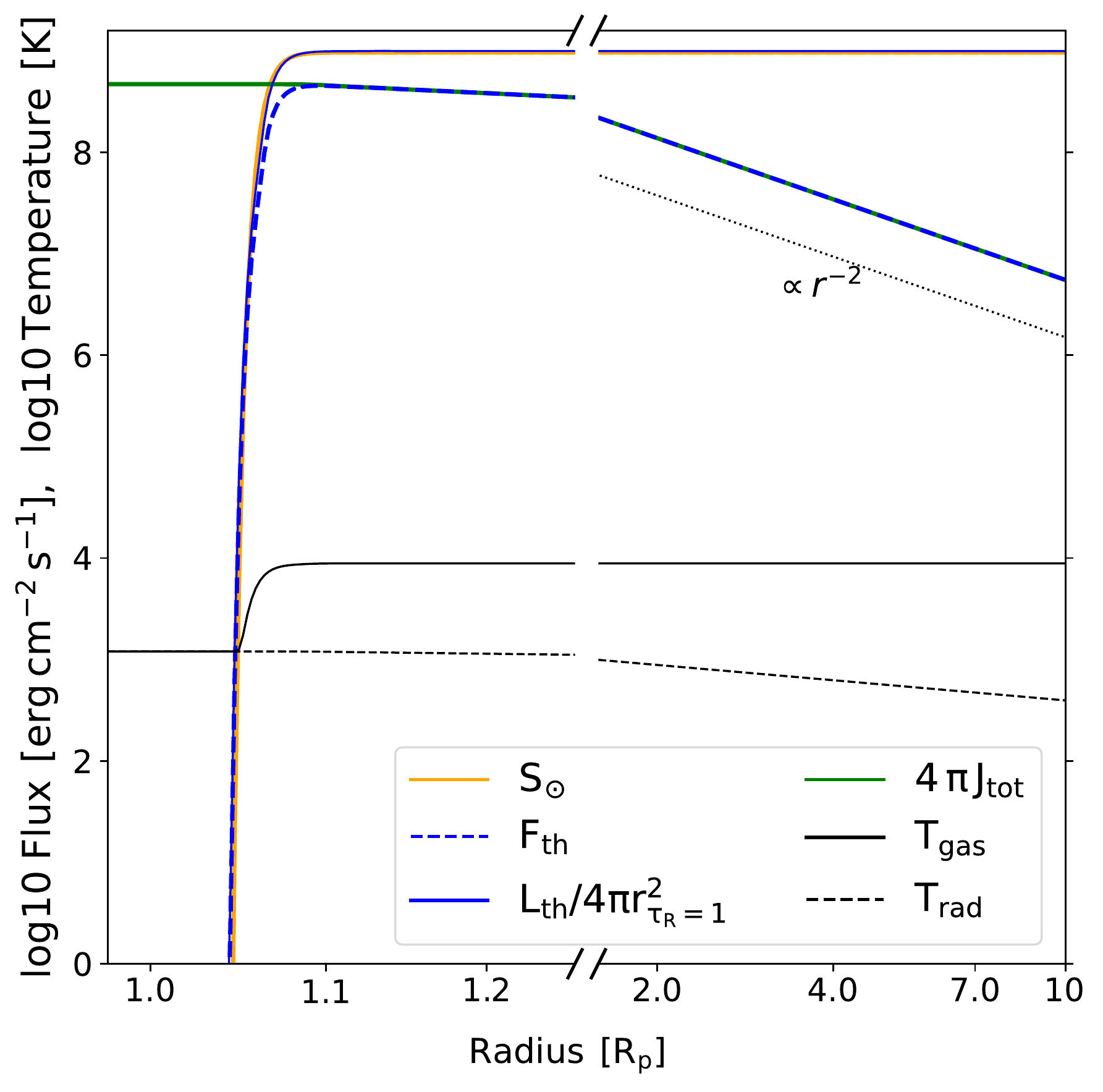}
\end{subfigure} 
\begin{subfigure}{0.32\textwidth} 
   \centering
   \includegraphics[width=1.0\textwidth]{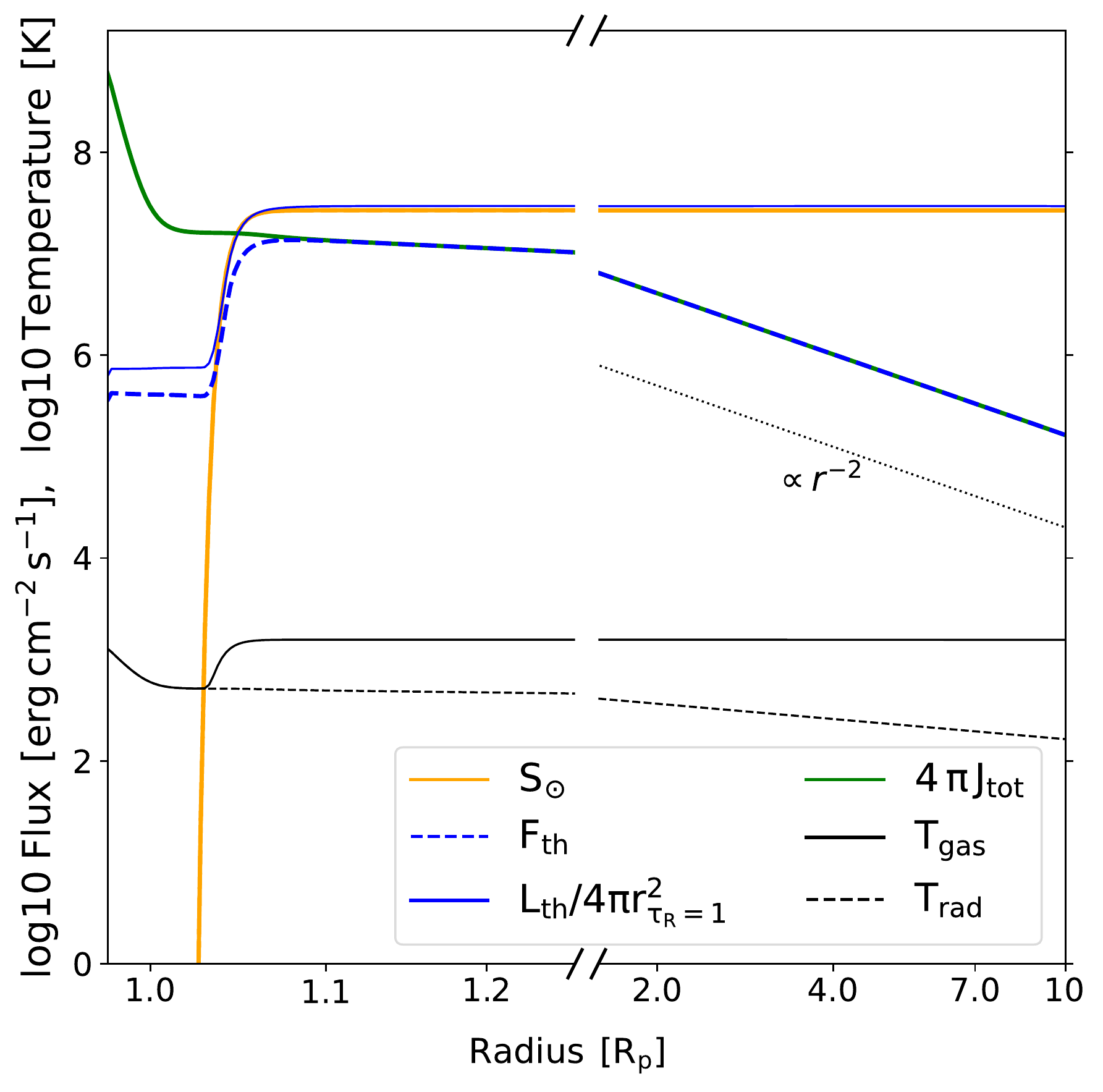}
\end{subfigure}

\caption{Same as Fig. \ref{fig:temperatures_analytic1}, but for irradiation for $\gamma>1$ without planetary luminosity (Red, \textbf{left} and \textbf{middle}) and with planetary luminosity (Green, \textbf{left} and \textbf{middle}). A notable difference with Fig. \ref{fig:temperatures_analytic1} is $T_{\rm gas}>T{\rm rad}$, due to the effects of inefficient re-radiation at $\gamma>1$ in Eqn. \ref{eq:guillot_multi}. } 
\label{fig:temperatures_analytic2}

\vspace*{+0.20cm}
\end{figure*}

\subsection{Single species, double-grey}
\label{sec:results_rad_dgrey}

After fixing $\xi=2$, we show a number of important atmospheric prototypes ranging from $\gamma<1$ to $\gamma>1$, which correspond to greenhouse and anti-greenhouse atmospheres. The planet in our simulation set to resemble HD~2094586b and its host star, i.e. a planet with $m_{\rm P} = 224 \,m_{\oplus}$, $R_{\rm P} = 1.35 R_{\rm Jup}$, with the initial atmospheric density profile constructed from a isothermal solution at $T_{\rm init} = T_{\rm eq} = 730$K. Our choice of $R_{\rm P}$, while taken from the transit radius literature, here means only the radius of the inner simulation boundary, at which reflective boundaries are applied.

\begin{figure*}
\hspace*{-0.5cm}
 \begin{subfigure}{0.32\textwidth} 
   \centering
   \includegraphics[width=1.0\textwidth]{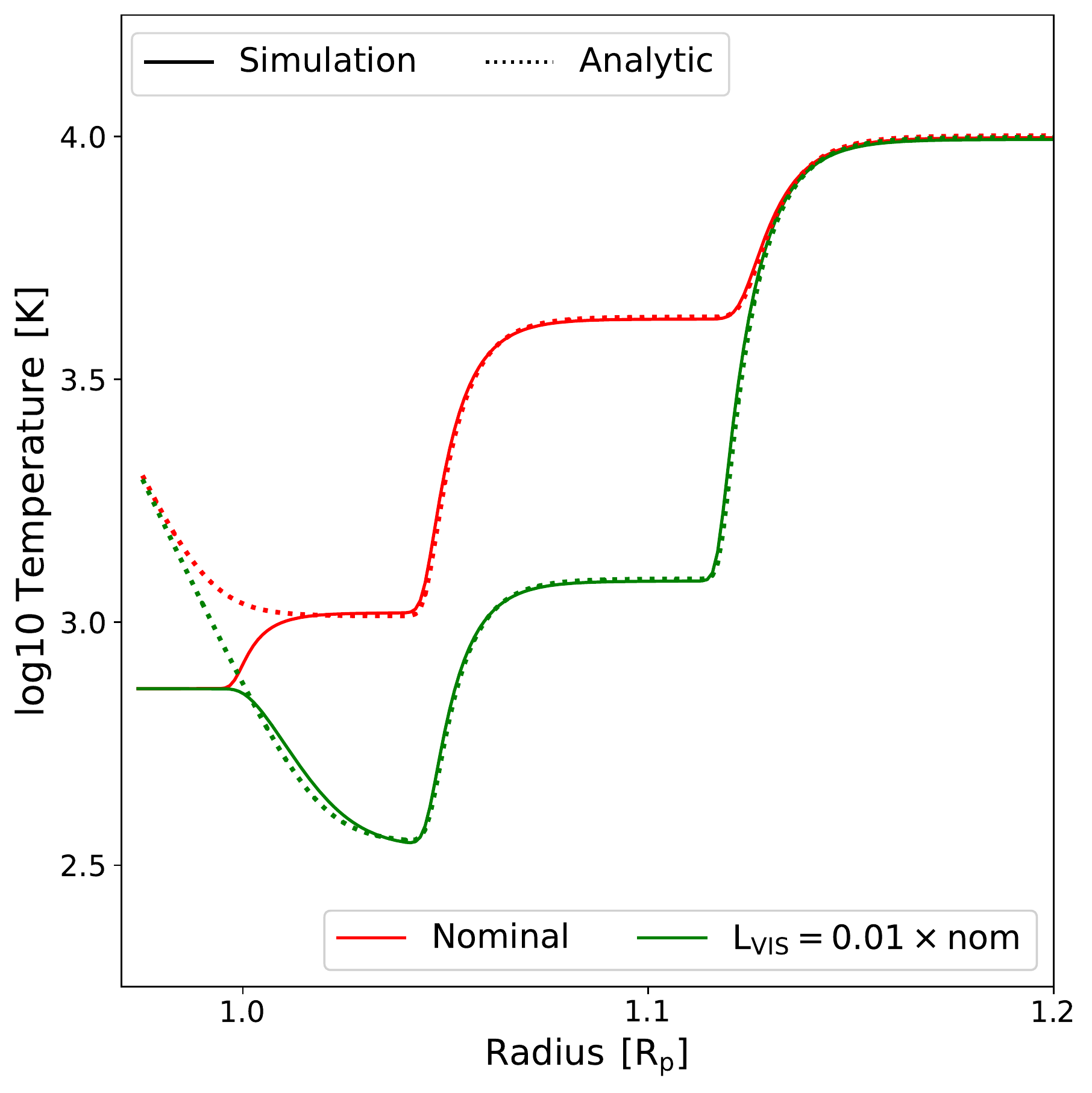}
\end{subfigure}%
\begin{subfigure}{0.34\textwidth} 
   \centering
   \includegraphics[width=1.0\textwidth]{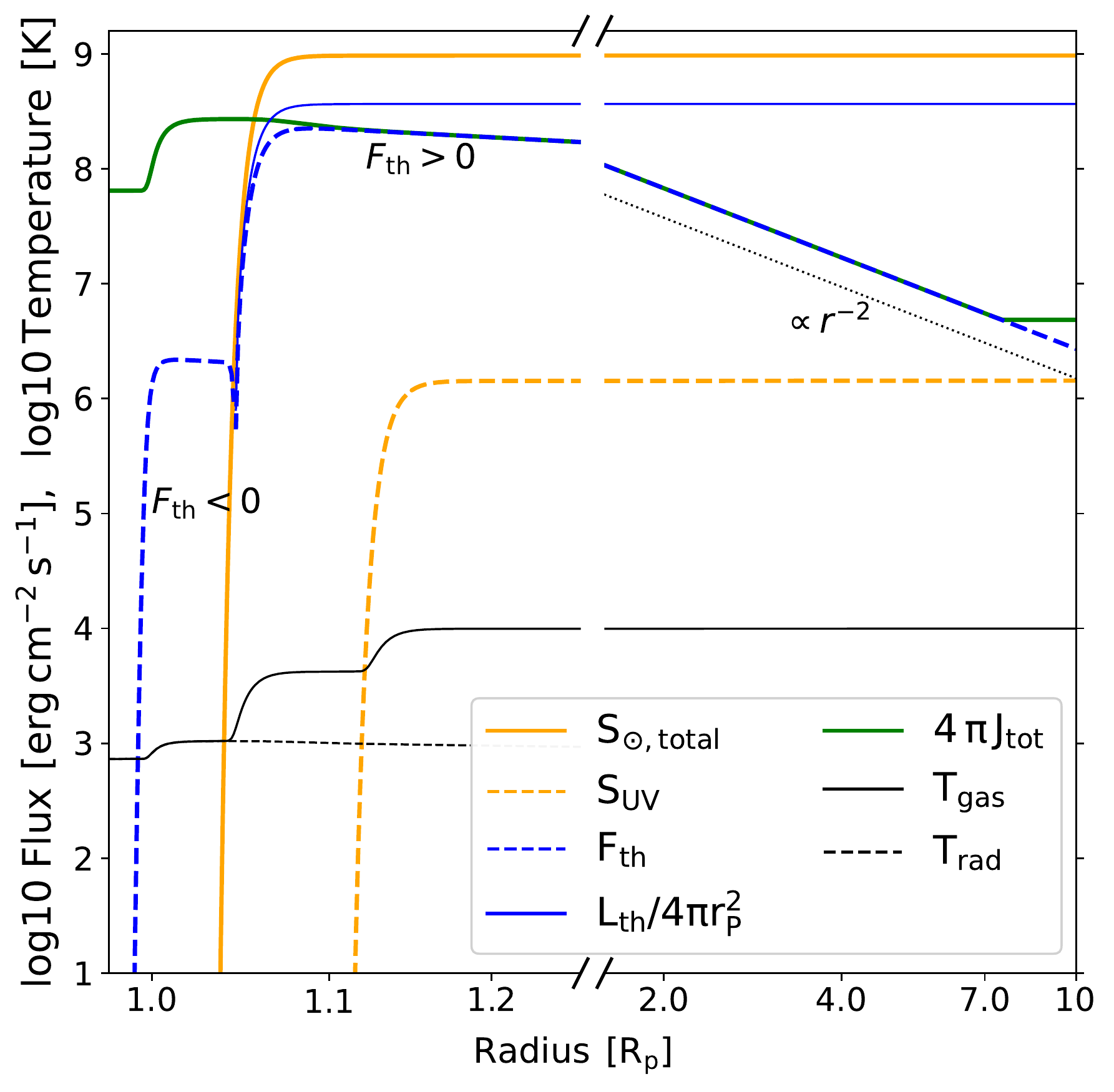}
\end{subfigure} 
\begin{subfigure}{0.34\textwidth} 
   \centering
   \includegraphics[width=1.0\textwidth]{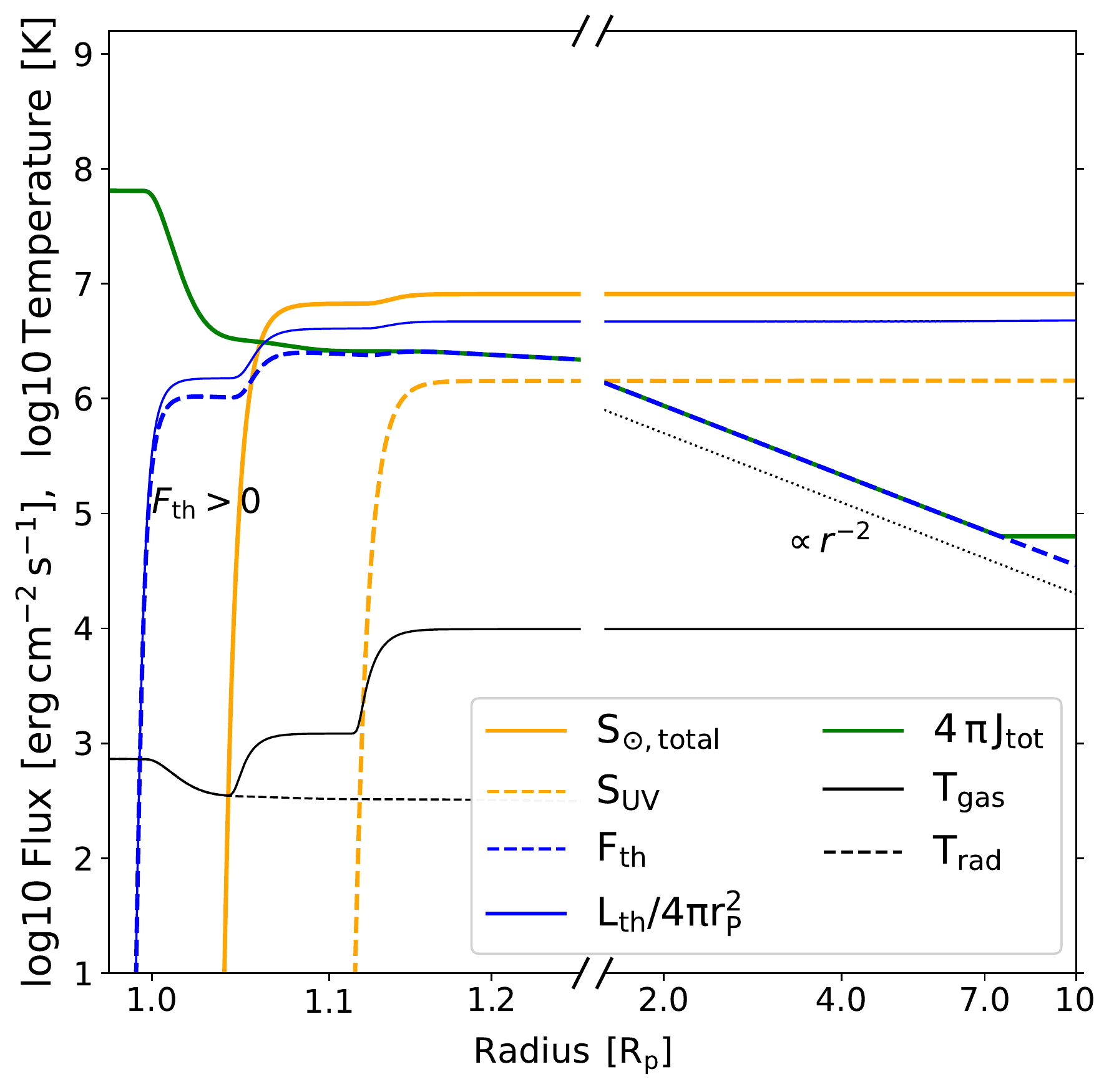}
\end{subfigure}

\caption{Same as Fig. \ref{fig:temperatures_analytic1}, but for a two-band irradiation solution, which is not in steady-state, after $t=10^6$s. This timescale can be comparable to the sound-crossing timescale of the simulation domain, hence it is important to note that an initial wind solution might not be the final steady state solution. Systematic deviations at low altitudes are due to our way of setting up the initial density-temperature structure. Below the $\tau_{\rm \odot,VIS}=1$ surface, the two cases differ significantly: The nominal case (Red, \textbf{left} and \textbf{middle}) re-emits absorbed sunlight further downwards ($F_{\rm th}<0$) and heats the atmosphere further, as there it is $\tau_{\rm R}<1$. The weakly irradiated case (Green, \textbf{left} and \textbf{right}) develops a cold stratosphere and the planet is not heated. The planetary thermal luminosity is slightly higher than the irradiated energy as our simulations do not have reached full thermal equilibrium with the deep atmosphere.} 
\label{fig:temperatures_analytic3}
\vspace*{+0.20cm}
\end{figure*}

The comparisons with Eqn. \ref{eq:guillot_multi} are shown in Figs. \ref{fig:temperatures_analytic1} and \ref{fig:temperatures_analytic2}. Skin temperatures and minimum temperatures as predicted by the preceeding description are shown as lower and upper dashed lines. All temperature gradients are correctly reproduced by the numerical solution. In the middle and right panel in Figs. \ref{fig:temperatures_analytic1}-\ref{fig:temperatures_analytic3} we plot important radiative quantities in order to check for code correctness and make it easier to understand the temperature profile. 

First, we show a case without irradiation (cyan curve) in order to test the diffusive regime of the flux-limited radiation transport. 
The final numerical profile shows an identical slope in the deep parts of the atmosphere, and a slight upwards deviation from the analytical profile until the plane parallel minimum temperature $T_{\rm min}$ is reached. From this point on, the temperature keeps decreasing, as flux density decreases via the $\vec \nabla \cdot \vec F$-term in Eqn. \ref{eqn:radiation_equation}, which has a $\frac{1}{r^2}$-dependence in spherical coordinates instead of being constant as in a plane-parallel geometry.

Next, we compare the profile of a irradiated, $\gamma$<$1$-atmosphere, which qualitatively shows similar behaviour to the internally heated case, with the same slope $\partial T/\partial \tau$. Also here does the temperature decrease below the nominal minimum temperature. Note how in the cooling-only case the planet is clearly the source of the atmospheric luminosity, whereas for the irradiation-only case the planet merely re-radiates the incoming flux after heating towards the steady-state temperature.

The case $\gamma$>$1$ with $T_{\rm int}=0$ is a typical case of a temperature inversion due to inefficient cooling, and shows radiative decoupling of $T_{\rm rad}$ and $T_{\rm gas}$. Finally, the case $\gamma$>$1$ with $T_{\rm int}=350 K$ shows inner and outer luminosity sources overlapping and raising the mid-altitude temperatures compared to the cooling-only case, although the absorption altitude is located much higher, at around $r=1.05 R_{\rm P}$.

\subsection{Single species, nonequilibrium multiband with grey opacities}
\label{sec:results_rad_multigrey}

Here, we set up a numerical test with two radiation bands and one species evaluated against a \cite{guillot2010}-like analytical solution. By assigning a low flux, but high irradiation opacity to one of the bands we can mimick the effects of a combined UV and bolometric irradiation scenario. The results can be seen in Fig. \ref{fig:temperatures_analytic3} for two different cases.

The planet is orbiting a G0 star with $R=1.2R_{\odot}$, $T_{\rm eff}=6070$K at a semimajor-axis distance of $0.05$ AU, emitting a perfect black-body spectrum. The resulting stellar bolometric flux at the planet is $S_{\odot,1} = 9.66 \times 10^{8}\rm\, erg\,cm^{-2}\,s^{-1}$ and the UV-flux is set to $S_{\odot,0} = 10^{6}\rm\, erg\,cm^{-2}\,s^{-1}$ and $T_{\rm int}=550$K. Without the photoionization module, the UV radiation simply acts as an additional bolometric heating source, depositing its full energy close to $\tau_{UV}=1$.
For reasons of simplicity, we set $\kappa_{\rm P}=\kappa_{\rm R}$, and the Planck-opacity ratios are set to constant values which mimick realistic values based on \citet{malygin2014} and \citet{Loth20} as $\gamma_{0} = 6\times10^6 $ and $\gamma_{1} = 3\times 10^2$ for both temperature curves. The results are shown in \ref{fig:temperatures_analytic3}.
The green curve differs from the red only in that it features a star with 100 times lower bolometric luminosity (while the UV luminosity remained constant), hence testing our numerical solution in the region closer to the planet, where the planetary internal luminosity plays a larger role.

The incoming solar and UV radiation are plotted as solid and dashed orange curves. Those radiation profiles are absorbed at altitudes corresponding to their fixed values of $\gamma_{b}$. The thermal re-radiation, i.e. re-radiation in band 1, $F_{1}\equiv F_{\rm th}$ is shown as dashed blue line. The UV re-radiation $F_{0}$ is omitted, as it is completely negligible in both the numerical solution and the analytic solution. In the optically thin parts of the atmosphere it is evident that the free-streaming limit is correctly reproduced, i.e. $F_{\rm th}=4\pi J_{th}$ and $F\propto r^{-2}$. 
The resulting planetary luminosity is $L = 4\pi r^2F_{\rm th}$ and should agree in steady-state with the incoming solar luminosity, when rescaled to the planetary surface, i.e. it should be $F_{\odot,\rm total} = L/(4\pi r_p^2)$. While the latter is close to being given in the case of the low-luminosity star, for the nominal case a larger deviation is evident. Here, a fraction of the luminosity is invested into heating the planet actively (the region marked with $F_{th} < 0$, widening the gap between incoming radiation and planetary luminosity.

In the lower atmosphere, i.e. the stratosphere at $R$<$1.05\, R_{\rm P}$, where $\tau_{\rm \odot,1}$>$1$ but $\tau$<$1$, the temperature is set by the condition $T=T_{\rm rad}$, where $T^4_{\rm rad} = J_{\rm tot}/(4\sigma_{\rm rad})$ is the radiation temperature of the total incoming solar flux. Even lower, at $R$<$1.00\, R_{\rm P}$, where it is $\tau$>$1$, the temperature in the analytic solution deviates. This is a result of our isothermal construction, as we initiate the temperature at a constant $730$K, and insufficient time has passed in the simulation to establish a diffusive temperature gradient in the optically thick inner regions. A diffusive gradient in the inner regions could be obtained by a long integration or a suitable choice of initial conditions, however. 


\section{Results - High-energy and ionization}
\label{sec:results_ion}
%
%

\begin{figure*}
\includegraphics[width=0.65\textwidth]{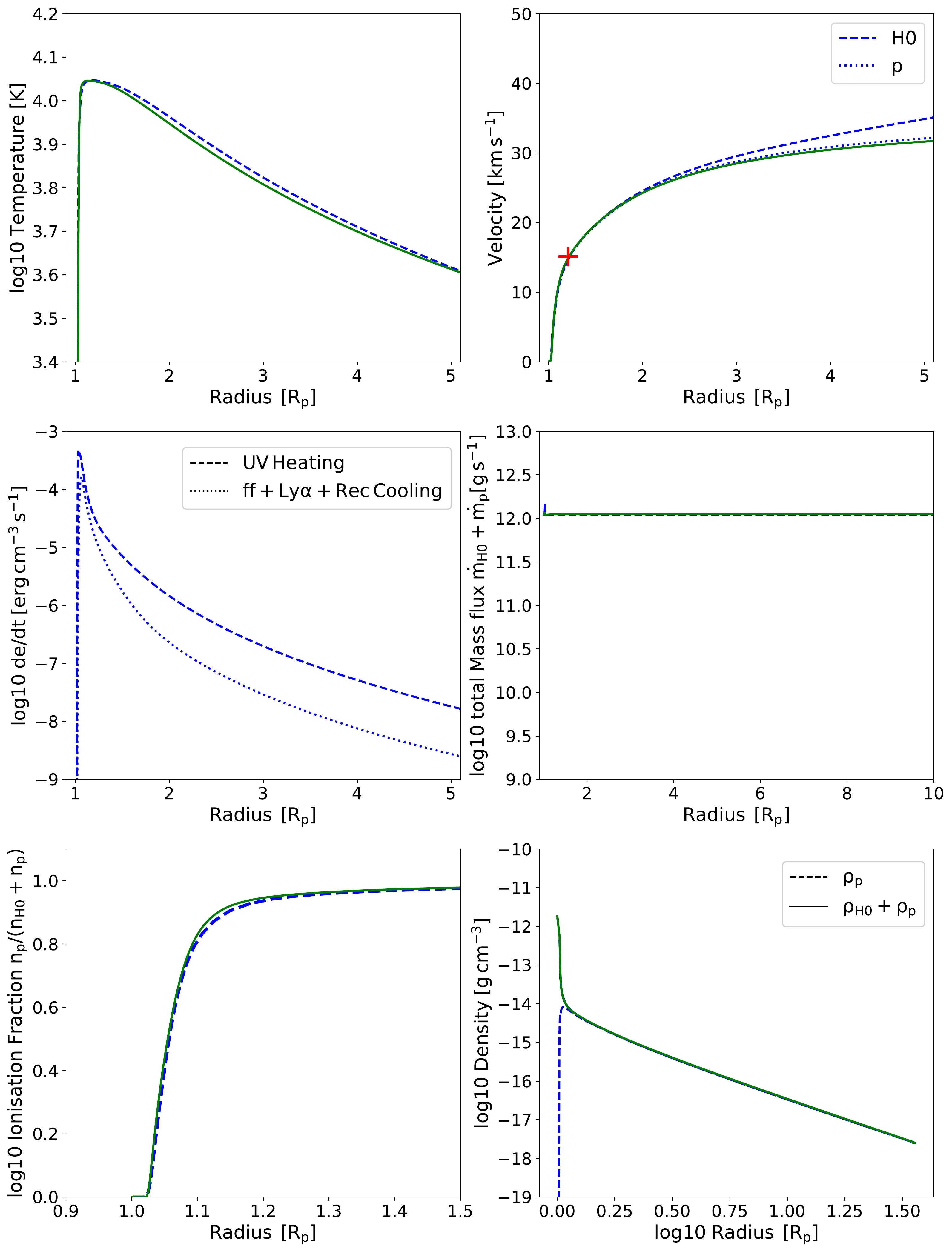}
\caption{Results from the ionization module test for GJ 436b for the high UV irradiation case ($ \rm S_{\odot,UV} =4\times 1.26\times 10^6 erg\,cm^{-2}\,s^{-1}$), which puts this outflow in the recombination-limited regime. Our simulation data shows the neutral hydrogen variables, unless otherwise specified and is shown as blue dashed or dotted lines, the canonical simulation is shown in green (see text for details). A constant recombination coefficient of $\rm \alpha=2.7\times10^{-13} cm^{-3}$ was used. The agreement between hydrodynamic variables is in general excellent, the escape rates only differ by $\sim 2\%$. The sonic point is noted as a red cross in the velocites \textbf{(Top Right)}. It is interesting to note that the protons, representing the majority of mass in this region, have a velocity that follows the canonical simulation, whereas the neutral hydrogen is about to decouple from the flow and accelerates outward, as the flow becomes more ionized and momentum keeps getting added to the neutral hydrogen according to Eqn. \ref{eq:ionization_momentum_correction_hydrogen}.  } 
\label{fig:c2ray_highenergy}
\vspace*{+0.20cm}
\end{figure*}

We now turn to show a test of the ionization module. The data we compare to is taken from  \cite{owenalvarez2016}), and generated with an implementation of $C^2$-Ray \citep{mellema2006, friedrich2012} in Zeus \citep{stonenorman1992}. The planet has $m_{\rm P}=21.25 m_{\oplus}$, a radius of $\rm r_p=2.756\times10^9\,cm$, an atmospheric mass in the simulation domain of $2\times 10^{-12}\, m_{\oplus}$ and an initial temperature of $T_{\rm eq}=712 \,K$, which we simultaneously use as lowest temperature floor, again identical to \citep{owenalvarez2016}.

This comparison requires the thermal re-radiation in our code to be inefficient, for which we simply multiply the radiation-matter coupling terms $\kappa_P \rho (J-B)$ in Eqns. \ref{eq:radtrans_jb_implicit} and \ref{eq:radtrans_fluxlimiter} by a factor of $10^{-100}$. This decouples the gas from the radiation field and hence only high-energy cooling terms come into effect, a situation identical to the physics in e.g. \cite{sekiya1980, watson1981, tian2005, MurrayClay2009, salz2016}. To reiterate, the equations solved are the ionization-equation, Eqn. \ref{eqn:ionization} along with the corresponding heating equations, Eqns. \ref{eq:ion_heating_h}-\ref{eq:ion_heating_total}. After the heating and cooling rates are computed and modified as a result of ionization, they are passed to the radiation solver, demonstrating its flexibility.

We then proceed to compare two limits identified as crucial in \cite{MurrayClay2009}, the advection-limited and recombination limited flows. Only the recombination-limited flow regime, i.e. the one occuring at high UV fluxes introduces new physics to our model, so we only plot this one. In both flux regimes our solutions show good agreement in the temperature, ionization rate, velocity and density variables, and excellent agreement down to better than 2$\%$ in the escape mass flux for the total escape rate of $\dot{m}_H + \dot{m}_{p^{+}}$.


\begin{figure*}
  \centering
   \includegraphics[width=0.90\textwidth]{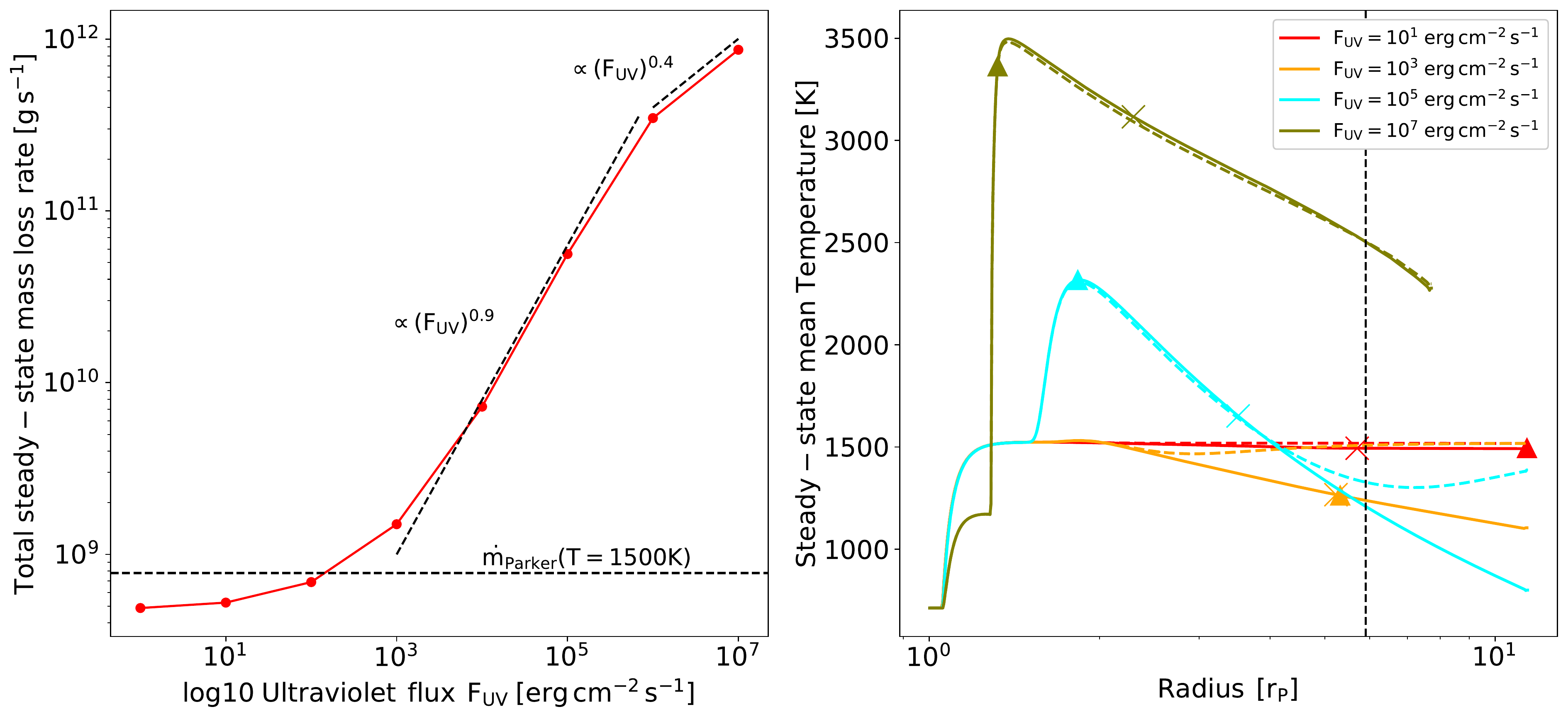}
\caption{Combined bolometric and photoevaporative models for GJ 436b. The mass loss rates (\textbf{Left}) show a transition from adiabatic nearly constant Parker-like core-powered mass-loss to energy-limited escape (seen as near-linear dependency on the UV flux) to recombination-limited mass-loss (seen as the high UV flux limit). The temperature profiles (\textbf{Right}) mark the sonic radius as crosses, the ionization front where $\rm n_H = n_{p+}$ as triangle, and the Hill-radius as vertical dashed line.} 
\label{fig:mdotvsuv}
\vspace*{+0.20cm}
\end{figure*}

\section{New applications}

We show some new applications and proof-of-concept simulations which are possible with our code, which we intend to follow-up with stand-alone publications.

\subsection{Thermally driven mass-loss and photoevaporation in the same simulation}

Here, we aim to combine two different physical processes which are discussed in the literature. Core-powered mass-loss, which is an isothermal mass-loss process at $T=T_{\rm eq}$, could in principle be supported by UV-driven mass-loss \citep{bean2021} achieving typically $T\sim 10^4 K$ in the upper atmosphere. However so far no simulations unifying those physical processes and their drivers over the entire spectral range exist.

For this set of simulations, we simulate the atmospheric escape of atomic hydrogen from the previous model of GJ436b, however now we include the bolometric radiation from the host M2-dwarf. We add a constant opacity to hydrogen in the bolometric band, in order to emulate absorption and efficient thermal heating by another species (which might be any molecular species, in a more complete simulation).

The setup consists of three species, $\rm (H,p^{+},e^{-})$, two incoming radiation bands with fluxes $(\rm F_{UV}, F_{Vis})$ where $\rm F_{Vis} = 1.1\times10^{7}\, erg\,cm^{-2}\,s^{-1}$, resulting from $\rm R_{star}=0.415 R_{\odot}$, $\rm T_{star}=3370K$  and one thermal band for the outgoing radiation $\rm F_{th}$. 
Opacities for $\rm H$ are $\rm ( \kappa_{UV}, \kappa_{Vis}, \kappa_{th}) = (1.1922\times 10^6, 1, 10) cm^2\, g^{-1}$. Thus, the visible radiation will be deposited in the lower atmosphere but at lower temperatures \citep{guillot2010}, compared to the UV radiation. For $\rm F_{UV} = 0$ there will be only limited $\rm H$ escape, as lower atmospheric temperatures are then insufficient to push the entire atmospheric mass out of the gravitational well. 

Atomic hydrogen starts in a hydrostatic state and the radiation is ramped up in the first $10^3$s to their maximum values in both bands. We show the results in steady state after $4\times10^6$s in steady-state, Fig. \ref{fig:mdotvsuv}.

It can be seen that at low UV fluxes the hydrogen escapes at a constant rate, close to the one given by the Parker wind. The simulation values are slightly lower due to the low-altitude atmosphere being at $T_{\rm eff}=712$K. This lowers the density at the sonic radius, and hence the escape rate, compared to the Parker wind. While the effect is only minor here, we think it should be worth investigating the importance of this decrease in escape rate in the context of core powered mass-loss. Photoionization becomes relevant at higher $\rm F_{UV}$, the upper atmospheric hydrogen temperature rises and the adiabatic (energy-limited) escape regime takes over, finally reaching the recombination-limited escape regime at very high $\rm F_{UV}$.

Because the solution for the temperature profile of irradiated planetary bodies presented by \cite{guillot2010} shows that the effects of solar energy deposited deep in the atmosphere and an internal core-luminosity are indistinguishable, we understand the hydrogen escape rates found here at $\rm F_{UV}=0$ as analoguous to core-powered mass-loss. However an implementation of a core-luminosity and a systematic examination of the process presented here in the future are necessary for clarity of distinguishing between differing approaches. Note that due to the additional thermal cooling which is also able to re-radiate thermalized UV photons, the escape rates and atmospheric temperatures are lower compared to the results shown in Fig. \ref{fig:c2ray_highenergy}.

This set of simulations uses the tidal potential from the star. As the potential minimum from star and planet is beyond the sonic point for our setup, the tidal potential is not the cause of the non-zero escape rate at $\rm F_{UV} \rightarrow 0$, it merely helps to speed up the finding of the steady state solution. As in the core-powered mass-loss scenario the thermal irradiation from the star is actually performing the heavy lifting, while the eponymous core luminosity changes the planetary radius from which material can be lifted off from, this simulation framework constitutes actual core-powered mass loss.

\subsection{Fully time-dependent, radiative simulation of gas giant formation}

We now demonstrate the capability of the code to form a giant planet in a radiative core-accretion simulation. Traditional approaches \citep{pollack1996, mordasini2012} use a stepping between steady-states in order to solve this problem, whereas we are capable of following the planets' formation in a fully time-dependent manner. 
In the core-accretion scenario, giant planets first accrete an envelope less massive than its core. This envelope slowly cools and contracts over a significant time, dubbed the hydrostatic phase. Once the first mass-doubling is reached, the contraction accelerates due to the action of the self-gravity of the gas. The accelerated contraction forces the gas to cool more intensely, which increases the accretion rate, again increasing the contraction rate. This process is called runaway gas-accretion. The runaway phase self-terminates in the classical literate once a disc-limited accretion rate is reached. In our case however, we do not impose a limit on the accretion rate. Instead, we imposed open boundary conditions which results in runaway accretion only slowing down when the simulation domain fills up with gas.

For the sake of simplicity, and to accelerate the formation of the gas giant, we use a seed core mass of 33$\rm m_{\oplus}$, akin to \cite{ayliffebate2012}, a constant opacity in the entire simulation domain of $\kappa=10^{-2}\,{\rm cm^2\,g}^{-1}$. In this calculation, we do not include the effects of convection, or hydrogen dissociation. We use a simulation domain size of $r_{\rm min} = 2\times10^{10}{\, \rm cm}$, $r_{\rm max} = 9\times10^{13}\,{\rm cm}$, $50$ simulation cells per decade, resulting in a total of $185$ cells. As our simulations are 1-D, the computation of self-gravity is achieved via the usage of the shell theorem, with the integrated mass $M(r) = 4\pi \int_0^{r} dr' r'^2 \rho(r')$ being used to re-compute the gravitational potential $\Phi(r) = GM(r)/r$ at every time-step.

The initial density profile is constructed using an adiabatic temperature structure, starting from a disc density at the simulation boundary of $\rho_0=3\times10^{-13}\,{\rm g\,cm}^{-3}$ at an initial entropy $\ln(p/\rho^{\gamma}) = 30.27$ with $\gamma=1.4$.
Our giant planet forms within 1600 years, reaching $\sim 8\rm M_{\rm Jup}$, at which point we terminate the simulation, while it reaches crossover mass after $\sim 600\rm \,yrs$. Those short timescales are found due to our very high choice of initial core mass, and the very low opacity. A first peak in the luminosity curve, at $\sim 100\,\rm yrs$ is due to the initial accretion of an envelope, until a hydrostatic solution is found that fills the planetary gravitational potential at given entropy. After the planet reaches a mass of $\sim 2 \rm M_{\rm Jup}$ at $900$ yrs, the simulation domain is filled up with quasi-hydroastatic gas again, however a high luminosity from ongoing contraction allows the planet to continue accreting. At this point the Hill-radius of the planet would be very large compared to the disc-scale height, and 3-D effects would need to be taken into account \citep{kley2001, tanigawa2012, judith2016, schulik2020}. Furthermore, it is known that the slow phase of quasi-hydrostatic growth can be lengthened by recycling the outer envelope entropy into the protoplanetary disc \citep{cimerman2017}, as well shortened by effects of dust growth \citep{movshovitz2010}, effects which to model is beyond the scope of this work.

\begin{figure}
  \centering
   \includegraphics[width=0.45\textwidth]{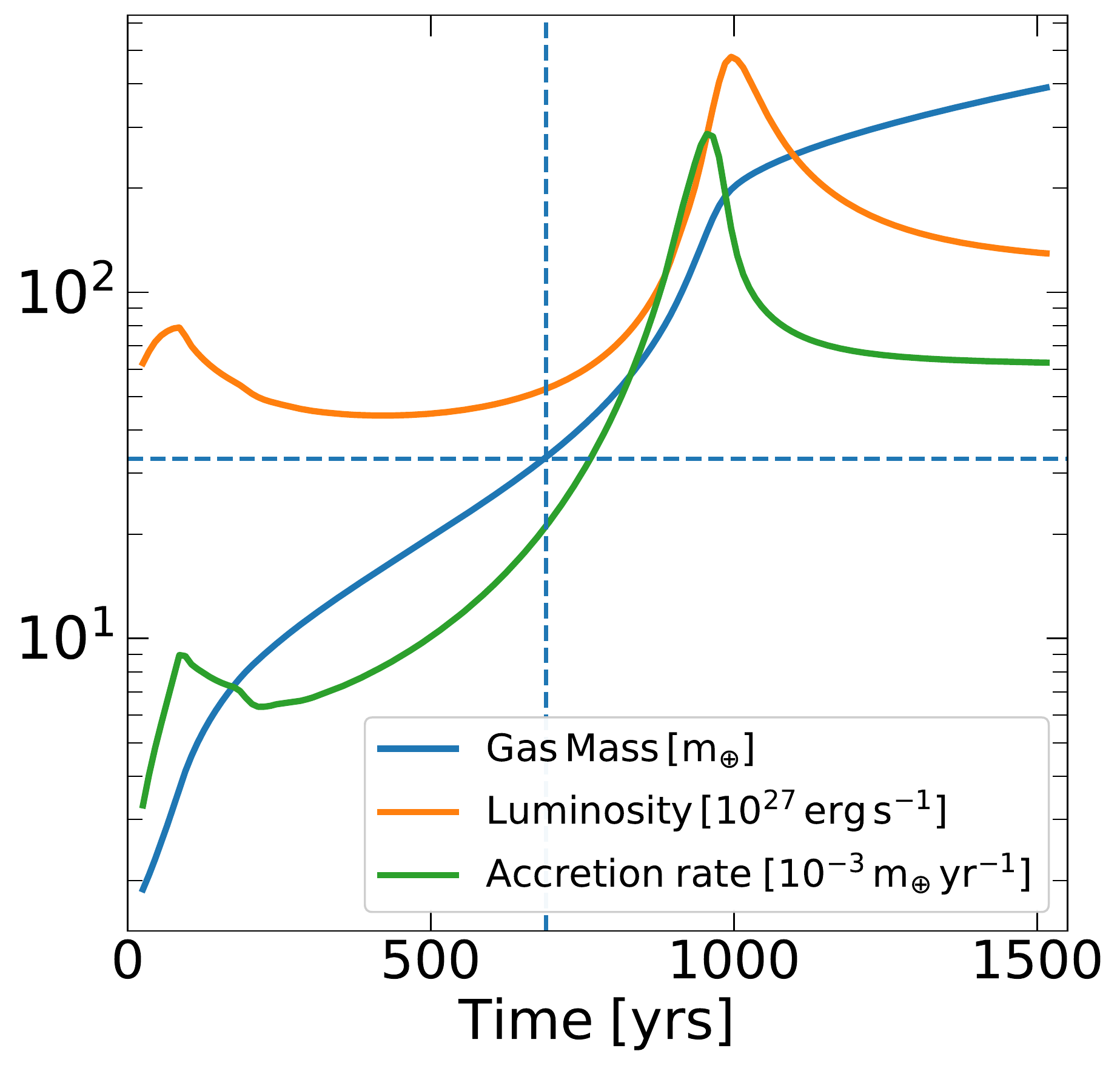}
\caption{Fully time-dependent giant planet model, including runaway gas accretion, starting with a cire mass of $\rm 33 m_{\oplus}$ and constant gas opacity of $\rm \kappa=10^{-2}\,cm^2\,g^{-1}$. The crossover-point in mass is marked with a horizontal and vertical dashed line, at around $700\,\rm yrs$, at which point gravitational self-contraction of the envelope becomes important, and the increase in luminosity starts supporting higher contraction rates, resulting in runaway gas-accretion and the formation of a $8.2 M_{\rm Jup}$ gas giant. } 
\label{fig:giantplanet}
\vspace{+0.20cm}
\end{figure}

\section{Future work and summary}

We have presented a new multi-species, multi-physics hydrodynamic code,  \textsc{Aiolos}, designed for simulating accretion onto, and hydrodynamic escape from, planetary atmospheres. We present a wide range of tests for the code. These include standard hydrodynamic and radiation hydrodynamic tests, along with a suite of problem-specific tests. We benchmark our code's results against existing models of atmospheric escape, such as EUV-driven mass-loss from hot Jupiters and the entrainment of trace species in an evaporative wind. For the first time, we demonstrate agreement between the atmospheric temperature structure computed by radiation transport computed in flux-limited diffusion approximation and analytic atmosphere models \citep{guillot2010}. Finally, we present new proof-of-concept calculations showing the transition between EUV-driven and thermally-driven mass loss (also known as core powered mass-loss, e.g. \citealt{gupta2019}). We also demonstrate the transition between between cooling-dominated and runaway gas accretion onto giant planets in a fully dynamical simulation.

\textsc{Aiolos} is already being applied to a range of problems. \citet{Booth2022} used \textsc{Aiolos} to model dust formation in the winds of ultra-short period planets.

Already in e.g. \cite{cooper2006} it was shown that non-equilibrium chemistry in the atmospheres of hot Jupiters can be very important, when the advective timescale is shorter than the kinetic timescale. For this reason, coupling with deep atmospheric codes, as well as a more advanced chemical solver seems like a necessary step to implement in the near-future.

In upcoming work we will furthermore explore the physics of full-spectrum atmospheric escape models with realistic opacities, which should prove to be a dramatic improvement over the simplistic results shown in Fig. \ref{fig:mdotvsuv}. Furthermore, the ionization of multiple species, such as H, C and O, and the transition from a molecular to atom-dominated atmosphere need to be addressed with an extension of the photochemistry scheme shown in this work. This work, once fulfilled, should provide valuable insight into the evolution and current state of the exoplanetary population.




\section*{Acknowledgements}
     We thank James Owen for thorough discussions and providing the ZEUS simulation data. RAB is supported by the Royal Society through University Research Fellowships. This project has received funding from the European Research Council (ERC) under the European Union’s Horizon 2020 research and innovation programme (Grant agreement No. 853022, PEVAP). M.S. would like to thank Shun Fai Ling for thorough proofreading. For the purpose of open access, the authors have applied a Creative Commons Attribution (CC-BY) licence to any Author Accepted Manuscript version arising. This project has recieved support from a 2020 Royal Society Enhancment Award.

\section*{Data Availability}
For the purpose of open access, the authors have applied a Creative Commons Attribution (CC-BY) licence to any Author Accepted Manuscript version arising. The code will be public and cleaned up once the manuscript is through the refereeing process.

%
%
\bibliographystyle{mnras}
\bibliography{bib}

\appendix
\section{Derivation of multi-band Guillot-type solutions}
\label{app:multi_guillot}

Here we provide a brief derivation of \autoref{eq:guillot_multi}, a multi-band generalization of \citet{guillot2010}'s approximate solution to the radiative transfer problem in plane-parallel atmospheres. Similar extensions already exist \citep[e.g.][]{parmentier2014}; however, we include our derivation since we retain some additional free parameters, such as $f_{\rm H}$ (defined below). We begin from the time-independent radiative-transfer equation for $J$ (\autoref{eqn:radiation_equation}), and consider only one band for the re-emitted radiation, but multiple bands for the stellar heating. By neglecting the hydrodynamic terms and assuming thermal equilibrium we may work with a single species without loss of generality. Together, these provide the starting point:
\begin{align}
    \deriv{H}{x} &= - \rho \kappa_{\rm P} \left[J - \frac{\sigma_{\rm SB} T^4}{\upi}\right] \label{eq:dH} \\
    H &= - \frac{1}{3 \rho \kappa_{\rm R}} \deriv{J}{x} \label{eq:H} \\
    0 &= \rho \kappa_{\rm P} \left[J - \frac{\sigma_{\rm SB} T^4}{\upi}\right]
        + f  \sum_b \frac{S_{\odot,b}}{4\upi} \exp(-\tau_b/\mu) \rho \kappa_{\odot, b}. \label{eq:TE}
\end{align}
Here we've explicitly kept the distinction between the Planck- and Rosseland-mean opacities in thermal bands. We have introduced $H=F/4\upi$ for ease of comparison with \citet{guillot2010} and also assumed the Eddington closure ($f_{\rm K}=1/3$ in the notation of \citealt{guillot2010}), denoted by $\mu$ the angle of the incident radiation, and introduced a factor $f$ that denotes how the stellar radiation is averaged over the planet. This factor should not be confused with $ f^{b}$ as used earlier in our numerical formulation, as in e.g. Eqn. \ref{eq:radtrans_Ts_implicit}, where it denotes the fraction of the Planck integral in band $b$. $f=1/4$ would be the usual isotropic average, which is used in \textsc{aiolos}, while $f=1$ would be the valid at the sub-stellar point ($\mu=1)$. 

Under the assumption that the various different opacities are constant, \autoref{eq:dH} and \autoref{eq:TE} may be solved together to give:
\begin{equation}
    H(\tau) = H(0) + f \sum_b \frac{S_{\odot,b}}{4\upi} \left[1 - \exp(-\tau \gamma_b /\mu)\right] \gamma_{\rm P} \mu,
\end{equation}
where we now explicitly denote $\tau$ as the Rosseland optical depth to thermal radiation, $\tau = \int \rho \kappa_{\rm R} {\rm d} x$ and introduced the factors $\gamma_b = \kappa_{\odot, b}/\kappa_{\rm R}$  and $\gamma_{\rm P} = \kappa_{\rm P}/\kappa_{\rm R}$. This choice of optical depth scale is arbitrary but follows \citet{parmentier2014}. Following \citet{guillot2010}, we associate the flux at large optical depth with the planet's internal luminosity, $H(\infty) = \sigma_{\rm SB} T_{\rm int}^4/4\upi$. Hence $H(0) = \sigma_{\rm SB} T_{\rm int}^4/4\upi - f \sum_b \frac{S_{\odot,b}}{4\upi}\gamma_{\rm P}\mu $.

We can now integrate \autoref{eq:H}, obtaining
\begin{align}
    J(\tau) &= J(0) + 3 H(\infty) \tau + 3 f \sum_b \frac{S_{\odot,b}}{4\upi} \frac{\mu^2}{\gamma_b}  \left[1 - \exp(-\tau \gamma_b /\mu)\right], 
\end{align}
where $\gamma_R = \kappa_{R}/\kappa_{\rm P}$. To close these equations we need an expression for $J(0)$. As in \citet{guillot2010}, we express $J(0)$ in terms of the Eddington factor $f_{\rm H} = H(0)/J(0)$. Together with \autoref{eq:TE}, this gives:
\begin{align}
    T^4 &= T_{\rm int}^4 \left[\frac{3}{4} \tau + \frac{1}{4 f_{\rm H}}\right]  \nonumber \\
        & +  \sum_b \frac{3 \mu}{4} f \frac{S_{\odot,b}}{\sigma_{\rm SB}} \left[\frac{1}{3f_{\rm H}} + \frac{\mu}{\gamma_b} + \left(\frac{1}{3\mu}\frac{\gamma_b}{\gamma_P} - \frac{\mu}{\gamma_b} \right) \exp(-\tau \gamma_b /\mu)\right].
\end{align}
In the main text we consider only $\kappa_{\rm R} = \kappa_{\rm P}$, i.e. $\gamma_{\rm P}=1$, and $\mu=1$. Finally, it is worth considering the appropriate choice of the Eddington factor, $f_{\rm H}$. \citet{guillot2010} recommends $f_{\rm H}=1/2$ since it agrees with more detailed solutions that take into account the anglular-dependence of the radiative-transfer equation. \textsc{Aiolos}, however, uses the flux-limited diffusion approximation (FLD) to solve the radiative-transfer problem. In FLD the flux in optically thin regions is limited by causality, i.e. radiation must not travel faster than the speed of light. This implies that $J=H$ in such regions and therefore we should expect $f_{\rm H} = 1$ to give better agreement with the numerical solutions computed with \textsc{aiolos}. $f_{\rm H}$ is only important around $\tau \sim 1$, however, so the differences can be expected to be limited to a small region.

\bsp	
\label{lastpage}
\end{document}